\documentclass[aps,pre,twocolumn,footinbib,superscriptaddress]{revtex4-2}

\usepackage[usenames, dvipsnames ]{xcolor}
\usepackage{amsmath, amsfonts, amssymb, amsthm, amstext, bm, bbm, multirow}
\usepackage[colorlinks=true, linkcolor=blue, citecolor=blue, 
urlcolor=blue, anchorcolor = blue, filecolor = blue, hypertexnames=true]{hyperref}
\usepackage{graphicx, caption, subcaption}
\usepackage{ragged2e}
\DeclareCaptionJustification{justified}{\justifying}
\usepackage{chngcntr}   

\newcommand{\bec}[1]{\mbox{\boldmath $ #1$}}
\newcommand{\Fig}[1]{Fig.~\ref{#1}}

\newcommand{\Eq}[1]{Eq.~(\ref{#1})}
\newcommand{\Eqs}[2]{Eqs.~(\ref{#1}) and~(\ref{#2})}

\newcommand{\mnras}{Mon. Not. R. Astron. Soc.}

\newcommand{\aap}{Astron. Astrophys.}

\newcommand{\physrep}{Phys. Rep.}

\begin{document}
	
	\title {Understanding   
		large-scale dynamos in unstratified rotating shear flows}
	
	\author{Tushar Mondal}
	\email{tushar.mondal@icts.res.in}
	\affiliation{International Centre for Theoretical Sciences, Tata Institute of Fundamental Research, Bengaluru 560089, India}
	
	\author{Pallavi Bhat}
	\email{pallavi.bhat@icts.res.in}
	\affiliation{International Centre for Theoretical Sciences, Tata Institute of Fundamental Research, Bengaluru 560089, India}
	
	\author{Fatima Ebrahimi}
	\email{ebrahimi@princeton.edu}
	\affiliation{Princeton Plasma Physics Laboratory, Princeton University, Princeton, NJ 08543, USA}
	\affiliation{Department of Astrophysical Sciences, Princeton University, Princeton, NJ 08544, USA}
	
	\author{Eric G. Blackman}
	\email{blackman@pas.rochester.edu}
	\affiliation{Department of Physics and Astronomy, University of Rochester, Rochester, NY 14627, USA}

	
	\begin{abstract}
		
		We combine simulations with new analyses that overcome previous pitfalls to explicate how non-helical mean-field dynamos grow and saturate in unstratified, magnetorotationally driven turbulence.  Shear of the mean radial magnetic field amplifies  the  azimuthal component.  Radial fields are regenerated by  velocity fluctuations that induce shear of radial  magnetic fluctuations, followed by Lorentz and Coriolis forces that source a negative off-diagonal component in the turbulent diffusivity tensor. We  present a simple schematic to illustrate this dynamo growth. A different part of the Lorentz force forms a third-order correlator in the mean electromotive force that saturates the dynamo. 
		
	\end{abstract}
	
	\maketitle
	
	
	
	Rotating shear flows are common in astrophysical accretion disks that drive phenomena such as planet formation, X-ray binaries and jets in protostars and compact objects. The disks are unstable to the magnetorotational instability (MRI), which drives turbulence that facilitates outward angular momentum transport \cite{1998RvMP...70....1B}. While MRI turbulence requires sufficiently coherent magnetic fields for sustenance \cite{2017MNRAS.472.2569B}, it simultaneously dissipates them via turbulent diffusion. Numerical simulations of MRI turbulence have consistently revealed the prevalence and importance of large-scale magnetic dynamos in characterizing and sustaining the steady-state \cite{1995ApJ...446..741B, 1996ApJ...464..690H, 2008A&A...488..451L, 2010MNRAS.405...41G, 2016MNRAS.462..818B, 2023PhRvE.108f5201M}. In mean-field dynamo theory \cite{1980opp..bookR....K, 2005PhR...417....1B, 2019JPlPh..85d2001R, 2021JFM...912P...1T}, the evolution of the large-scale magnetic field is governed by:
	\begin{equation}
		\frac{\partial \bar{\bm{B}}}{\partial t} = \bm \nabla \times (\bar{\bm{U}} \times \bar{\bm{B}} + \bar{\bm{\mathcal{E}}} - \eta \mu_0 \bar{\bm{J}}) , \label{eq:meanB}
	\end{equation}
	where $\bar{\bm{U}}$, $\bar{\bm{B}}$, and $\bar{\bm{J}}$ represent the large-scale (mean) velocity, magnetic field, and current density, respectively. The mean electromotive force (EMF), $\bar{\bm{\mathcal{E}}} = \overline{\bm u \times \bm b}$, encapsulates the correlation between fluctuating velocity and magnetic fields. In traditional mean-field closure, the EMF is modeled as $\mathcal{\bar E}_i = \alpha_{ij} \bar B_j + \beta_{ijk} \bar B_{j,k} = \alpha_{ij} \bar B_j - \eta_{ij} \bar J_j$, where $\alpha_{ij}$ and $\eta_{ij}$ (or $\beta_{ijk}$) are turbulent transport coefficients determined by small-scale dynamics.

	Determining the physical origin of the coefficients in this formalism that best capture large-scale MRI growth in simulations has been an active area of research. Several theoretical investigations, including studies of nonlinear mode coupling in the subcritical nature of the MRI dynamo \cite{2007PhRvL..98y4502R, 2011PhRvE..84c6321H, 2017A&A...598A..87R}, nonlinear transverse cascades \cite{2020ApJ...904...47M, 2024MNRAS.530.2232H}, the role of turbulent resistivity on the self-regulation of MRI turbulence \cite{2023MNRAS.521.5952B, 2024MNRAS.534.3144B}, and minimal ingredients required for large-scale field growth from quasi-linear analysis \cite{ebrahimi2009saturation, 2011PhRvL.107y5004H, 2016MNRAS.459.1422E}, have been conducted to better understand MRI turbulence and associated dynamo behavior.

	In vertically stratified, differentially rotating disks, the traditional $\alpha$--$\Omega$ dynamo possibly explains large-scale magnetic field generation: the $\alpha$-effect, driven by helical turbulence, generates poloidal fields, while the $\Omega$-effect, from differential rotation, shears them into toroidal fields \cite{2010MNRAS.405...41G, 2015ApJ...810...59G, 2016MNRAS.456.2273S, 2024MNRAS.530.2778D}. However, simulations of MRI turbulence in unstratified rotating shear flows exhibit predominantly non-helical, large-scale magnetic fields \cite{2016MNRAS.456.2273S, 2016MNRAS.462..818B, 2022A&A...659A..91W, 2022MNRAS.517.2639Z, Skoutnev+2022}, raising fundamental questions about alternative dynamo mechanisms that do not rely on mean helicities.

	A leading hypothesis attributes such non-helical large-scale dynamos to a negative off-diagonal component of the turbulent diffusivity tensor, arising from shear, rotation, or their combination. Rotation alone produces the $\bm{\Omega} \times \bm{J}$ (R\"{a}dler) effect \cite{1986AN....307...89R}, while shear may induce the (kinetic or magnetic) shear-current effect \cite{2003PhRvE..68c6301R, 2004PhRvE..70d6310R, 2015PhRvL.115q5003S}. However, the role and reliability of these effects remain debated, as their interpretation is limited by closure approximations and difficulties in simultaneously extracting all turbulent transport coefficients amid strong fluctuations and oscillatory large-scale fields \cite{2021MNRAS.507.5732Z}.

	Another pressing question is how dynamos driven by negative $\eta_{yx}$ saturate. Previous studies hypothesized that saturation occurs through a sign reversal of $\eta_{yx}$ after the initial growth phase \cite{2016JPlPh..82b5301S}. However, Ref.~\cite{2016MNRAS.456.2273S} showed that $\eta_{yx}$ remains consistently negative throughout both the linear and nonlinear stages. A complete physical understanding of non-helical MRI large-scale dynamos and their saturation remains elusive.

	To address these outstanding issues, we present a fundamentally new approach based on a self-consistent formulation of the mean EMF, whose mathematical framework was first introduced through direct statistical simulations (DSS) \cite{2023PhRvE.108f5201M}. The EMF is constructed from the evolution equations of the Faraday tensor $\bar{F}_{ij} = \overline{u_i b_j}$, which contains interaction terms involving the Coriolis force and background shear---core features of rotating shear flows. These terms mediate couplings between different components of $\bar{F}_{ij}$, and by systematically analyzing these couplings, we derive exact expressions for the EMF and associated turbulent transport coefficients. The resulting EMF contains not only the familiar terms proportional to $\bar B_i$ and its gradients, as in traditional mean-field closures, but also contributions from gradients of $\bar U_i$, third-order correlators, and pressure fluctuations. In earlier DSS-based approaches, closure models were required for the third-order terms \cite{2023PhRvE.108f5201M}. Here, we perform direct numerical simulations (DNS) using the \textsc{Pencil Code} \cite{2021JOSS....6.2807P} to compute each EMF contribution explicitly, avoiding any a priori closure. Unlike previous methods, 
	our formulation yields explicit, self-consistent expressions without relying on fitting procedures or closure approximations. This enables us to unambiguously identify the dominant source term responsible for large-scale magnetic field generation. To uncover its physical origin, we further analyze the evolution equations of the relevant fluctuating fields that make up the correlators. 
	We refer to the growth mechanism as the rotation-shear-current (RSC) effect.
	It operates via a negative turbulent diffusivity component, $\eta_{yx}$, which remains  negative throughout both the linear and nonlinear stages. 
	
\begin{figure*}
		\centering
		
		\begin{minipage}[t]{0.33\textwidth}
			\includegraphics[width=0.99\columnwidth]{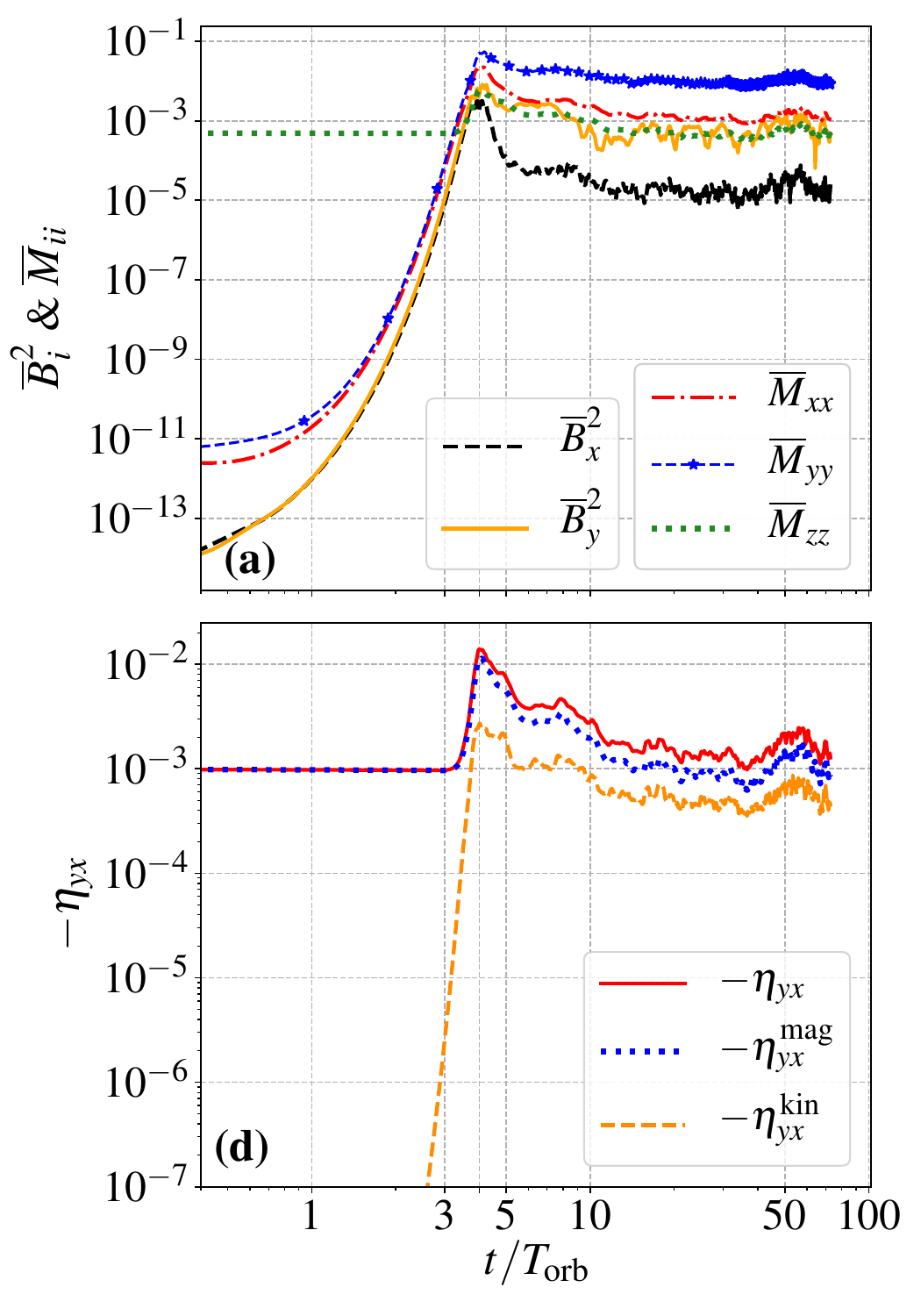}
		\end{minipage}%
		\hfill
		\begin{minipage}[t]{0.665\textwidth}
			\raisebox{90pt}{%
				\begin{minipage}[t]{\textwidth}
					\includegraphics[width=0.5\textwidth]{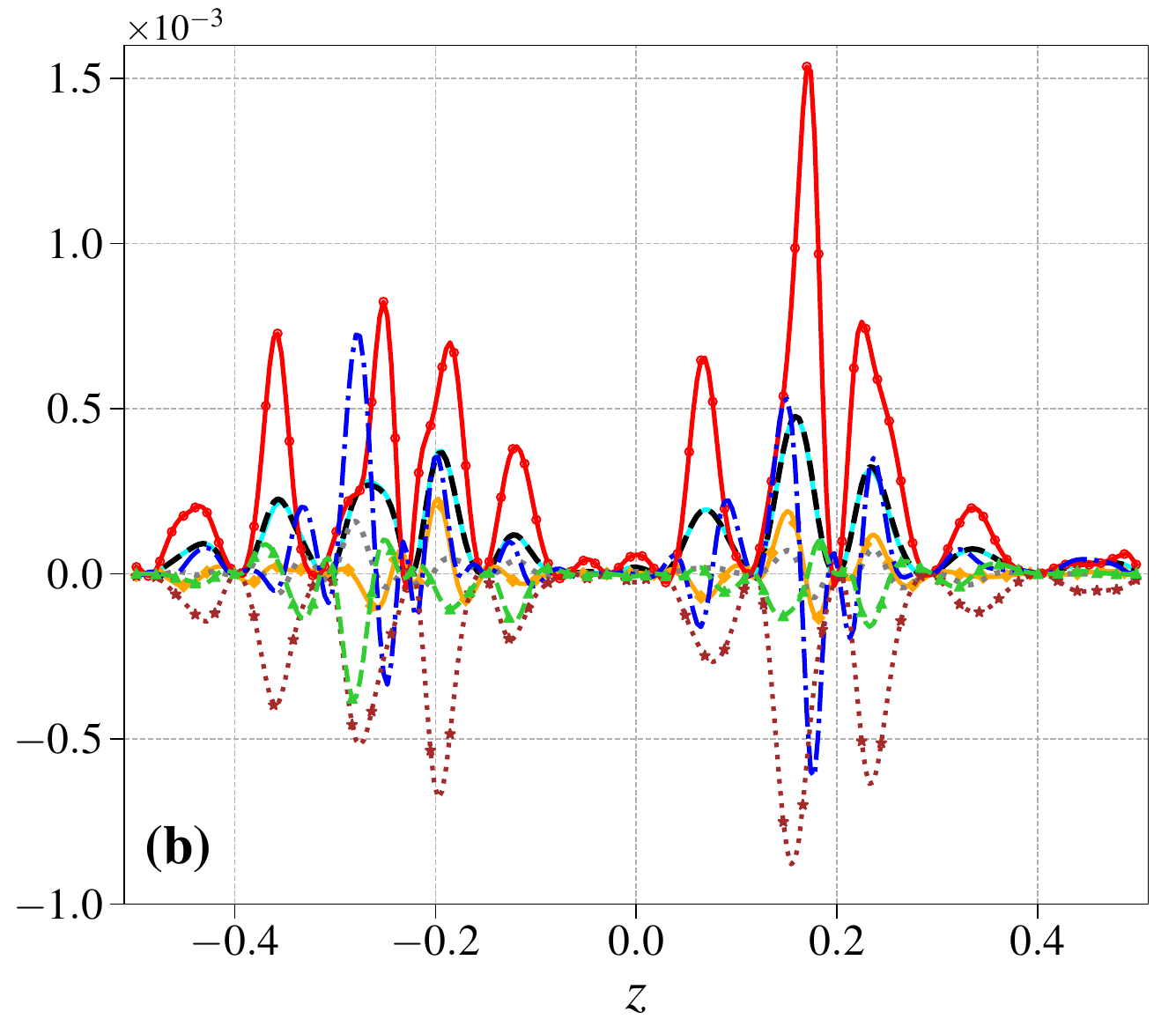}%
					\hfill
					\includegraphics[width=0.5\textwidth]{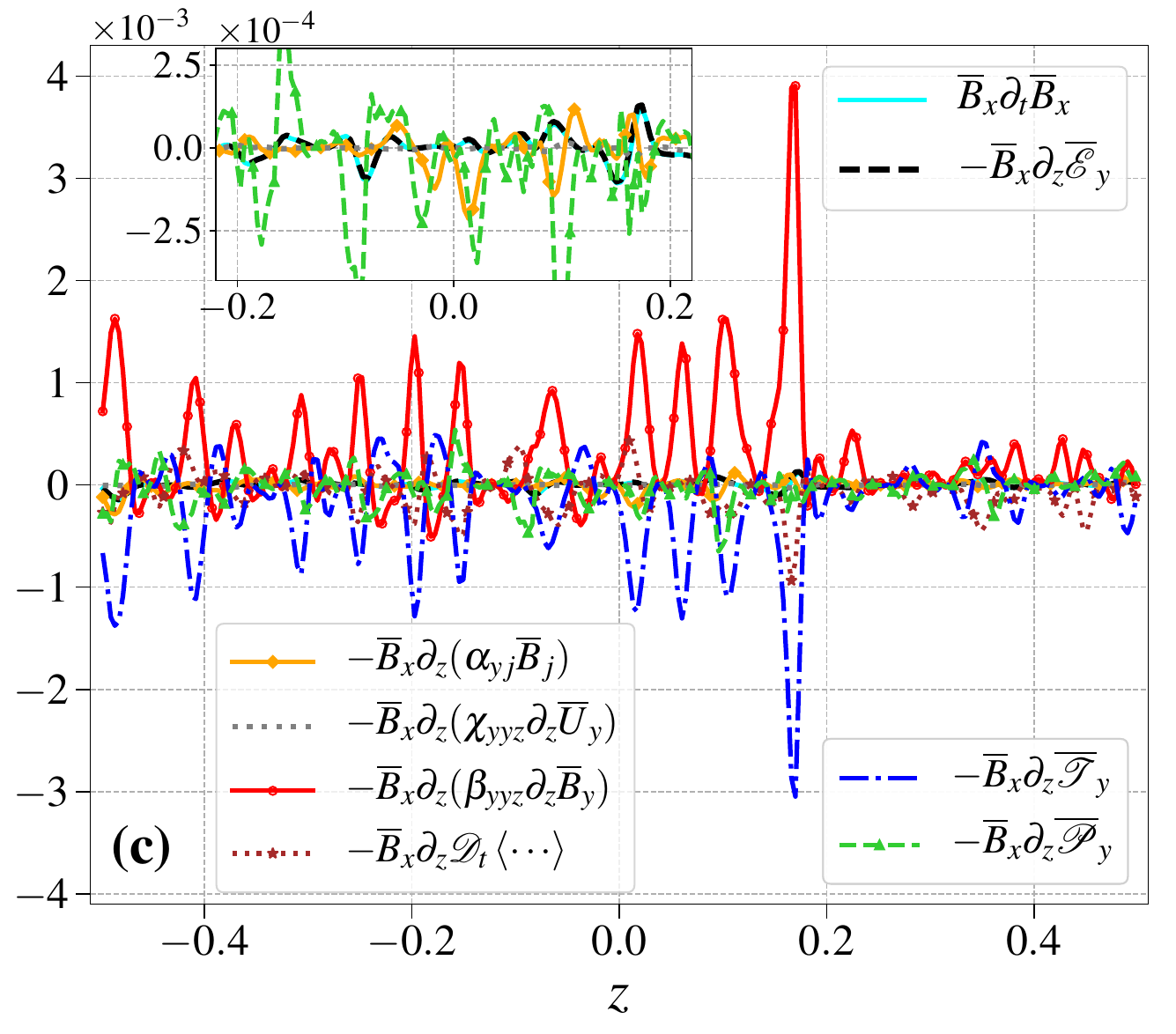}
					\caption{
						(a) Time evolution of magnetic energy densities (scaled by two) for $xy$-averaged large-scale and fluctuating fields.
						(b), (c) Spatial variation of individual terms from the vertical gradient of $\mathcal{\bar E}_y$ (Eq.~\ref{eq:meanBx_xy_2}) responsible for generating $\bar B_x(z)$, shown during (b) exponential growth ($t/T_{\text{orb}} = 3.34$) and (c) nonlinear stage ($t/T_{\text{orb}} = 60$). Line styles and colors are consistent in (b) and (c); legend is shown in panel (c).
						(d) Time evolution of the off-diagonal turbulent diffusivity $\eta_{yx}$ (Eq.~\ref{eq:eta_yx}). In (a) and (d), additional averaging along $z$ is applied.
					}
					\label{fig:BxdtBx_z_xyaver}
				\end{minipage}
			}
		\end{minipage}
		
	\end{figure*}

	Horizontal planar averaging, denoted by $ \overline{\cdot}$ or $\left< \cdot \right>$, defines the large-scale field in our investigation of large-scale dynamos in MRI turbulence. Here we adopt Cartesian coordinates $(x,y,z)$ corresponding to the radial, azimuthal, and vertical directions of the disk, respectively. \Fig{fig:BxdtBx_z_xyaver}(a) shows the time evolution of magnetic energy densities (scaled by two) for large-scale ($\bar B_i^2$) and fluctuating ($\bar M_{ii} = \overline{b_i^2}$) components. Fluctuating fields are comparable to or even stronger than large-scale fields during the exponential growth phase, with the azimuthal component dominating at both scales throughout nonlinear saturation. To understand the origin and sustenance of these large-scale fields, we analyze the $xy$-averaged mean-field induction equations derived from Eq.(\ref{eq:meanB}):
	\begin{subequations}
		\begin{align}
			\partial_t \bar B_x &= - \partial_z \mathcal{\bar E}_y - \bar U_z\partial_z\bar B_x + \bar B_z\partial_z\bar U_x ,
			\label{eq:meanBx_xy} \\
			\partial_t \bar B_y &= - q\Omega \bar B_x + \partial_z \mathcal{\bar E}_x - \bar U_z\partial_z\bar B_y + \bar B_z\partial_z\bar U_y .
			\label{eq:meanBy_xy}
		\end{align}
		\label{eq:meanB_xy}
	\end{subequations} 
	Here, $-q\Omega \bar B_x$ is the shear term due to a linear background shear, $\bec{U}^0 = -q \Omega x \hat{y}$, with $q=3/2$ for Keplerian disks. The terms $\bec{\bar U}\cdot \nabla \bec{\bar B}$ and $\bec{\bar B}\cdot \nabla \bec{\bar U}$ represent advection and stretching, respectively. The mean EMF components are $\mathcal{\bar E}_x = (\bar F_{yz} - \bar F_{zy})$ and $\mathcal{\bar E}_y = (\bar F_{zx} - \bar F_{xz})$. As established in Ref.~\cite{2023PhRvE.108f5201M}, $\bar B_x$ is generated by the vertical gradient of the azimuthal EMF, $\partial_z \mathcal{\bar E}_y$, while shear stretches $\bar B_x$ into $\bar B_y$ via the $\Omega$-effect. In contrast, the vertical gradient of the radial EMF, $\partial_z \mathcal{\bar E}_x$, extracts energy from $\bar B_y$, acting as a sink. The key is to understand how $\mathcal{\bar E}_y$ drives the growth of $\bar B_x$ in a self-sustaining dynamo.


	The EMF can be constructed from the evolution equations of the Faraday tensors (see Appendix~\ref{sec:appendix_EMF}):
	\begin{align}
		\mathcal{\bar E}_y = & \frac{-1}{q(2-q)\Omega} \Bigg[ \Big\{- q \mathcal{D}_t \bar F_{yz} + (2-q) \mathcal{D}_t \bar F_{zy} \Big\}       \nonumber\\
		& - 2\bar F_{zz} \partial_z \bar U_y         
		+ \frac{1}{\rho} \Big\{q\bar M_{zz} + (2-q)\bar R_{zz} \Big\} \partial_z \bar B_y    \nonumber\\
		& + \bar B_k \Bigg\{ q \left(\langle u_y \partial_k u_z \rangle + \frac{\langle b_z \partial_k b_y \rangle}{\mu_0 \rho} \right) - (2-q) \nonumber\\
		& \hspace{0.75 cm}  \times \left(\langle u_z \partial_k u_y \rangle + \frac{\langle b_y \partial_k b_z \rangle}{\mu_0 \rho} \right) \Bigg\} \Bigg] + \mathcal{\bar P}_y + \mathcal{\bar T}_{y}.
		\label{eq:emfy}
	\end{align}
	Here, $\mathcal{D}_t =  \partial_t - q\Omega x \partial_y$ includes advection by the background shear. The third-order and pressure fluctuation terms are 
	$\mathcal{\bar T}_{y} = - \left\{ q \mathcal{\bar T}^F_{yz} - (2-q) \mathcal{\bar T}^F_{zy} \right\}/ q(2-q)\Omega $, and 
	$\mathcal{\bar P}_y = \left\{ q \langle b_z\partial_y p_t \rangle - (2-q) \langle b_y\partial_z p_t \rangle \right\} / q(2-q)\rho \Omega $, 
	where $\mathcal{\bar T}^F_{ij} = \left\langle u_i b_k \partial_k u_j + b_j b_k \partial_k b_i - u_k \partial_k F_{ij} \right\rangle$, and $p_t = p_g + p_b$ is the total (gas + magnetic) pressure fluctuation.
	The EMF $\mathcal{\bar E}_y$ consists of terms proportional to $\bar B_i$, its gradients, gradients of $\bar U_i$, third-order correlators, pressure fluctuations, and time derivatives of $\bar F_{ij}$. 
	Substituting \Eq{eq:emfy} into \Eq{eq:meanBx_xy}, we obtain:
	\begin{align}
		\partial_t \bar B_x = - \partial_z \mathcal{\bar E}_y =  
		- \partial_z \Big[& \alpha_{yj} \bar B_j + \beta_{yyz} \partial_z \bar B_y + \chi_{yyz} \partial_z \bar U_y \nonumber\\ 
		& + \mathcal{\bar P}_y + \mathcal{\bar T}_y + \mathcal{D}_t \langle \cdots \rangle \Big],
		\label{eq:meanBx_xy_2}
	\end{align}
	where $\alpha_{ij}$, $\beta_{ijk}$, and $\chi_{ijk}$ are the turbulent transport coefficients, derived from \Eq{eq:emfy}. The off-diagonal turbulent diffusivity component is given by
	\begin{equation}
		\eta_{yx} = \beta_{yyz} = - \frac{1}{\rho \Omega}\left[\frac{1}{2-q} \bar M_{zz} + \frac{1}{q} \bar R_{zz} \right].
		\label{eq:eta_yx}
	\end{equation}
	For Keplerian shear ($q = 1.5$), $\eta_{yx}$ remains negative because $\bar M_{zz}$ and $\bar R_{zz}$---the energies associated with vertical magnetic and velocity fluctuations---are always positive, thereby driving the amplification of large-scale magnetic fields.
	The magnetic–shear–current (MSC) effect \cite{2016JPlPh..82b5301S} also originates from a negative $\eta_{yx}$, but the RSC effect in \Eq{eq:eta_yx} is fundamentally distinct from the MSC effect. In the MSC framework, a negative $\eta_{yx}$ arises from magnetic fluctuations interacting with shear and a large-scale magnetic field gradient, with the pressure response of velocity fluctuations playing a central role. Rotation is not essential in this process, and the turbulence is maintained by external forcing.
		In contrast, in our simulations the turbulence is entirely MRI-driven, without external forcing, and is therefore intrinsically tied to rotation. We show later in this work that the Coriolis force plays a central role in coupling Lorentz-force–driven fluctuations to the EMF.
		In the absence of rotation, the magnetic term in Eq.~(\ref{eq:eta_yx}) changes sign and becomes diffusive, so $\eta_{yx}$ no longer guarantees a net dynamo effect. The overall sign may then depend on the energy spectra \citep{2021MNRAS.507.5732Z}.

	Figs.~\ref{fig:BxdtBx_z_xyaver}(b) and (c) show how each term in $\mathcal{\bar E}_y$ contributes to the generation of $\bar B_x(z)$ during two key MRI stages: (b) the exponential growth phase at $t/T_{\text{orb}} = 3.34$, and (c) the nonlinear saturation stage at $t/T_{\text{orb}} = 60$. For clarity, we multiply $\bar B_x$ on both sides of \Eq{eq:meanBx_xy_2}, ensuring the source term remains positive regardless of the sign of $\bar B_x$ growth.
	During the exponential growth (\Fig{fig:BxdtBx_z_xyaver}b), the dominant source is the magnetic field gradient term (linked to $\beta_{yyz}$ or $\eta_{yx}$), which amplifies $\bar B_x(z)$. The time-derivative term is predominantly dissipative, while terms proportional to $\bar B_i$ and velocity gradients are negligible. The third-order correlator exhibits localized variations that can either enhance or counteract the mean-field growth.
	In the nonlinear stage (\Fig{fig:BxdtBx_z_xyaver}c), third-order correlators become dominant and counteract the $\eta_{yx}$-driven growth, sustaining dynamo self-regulation.

	The large-scale dynamo mechanism, governed by $\eta_{yx} \bar J_x$ or $\beta_{yyz} \partial_z \bar B_y$, is known as the rotation-shear-current effect \cite{2023PhRvE.108f5201M}. According to \Eq{eq:eta_yx}, the total off-diagonal diffusivity, $\eta_{yx} = \eta_{yx}^{\text{mag}} + \eta_{yx}^{\text{kin}}$, comprises a magnetic contribution (linked to $\bar M_{zz}$) and a kinetic one (linked to $\bar R_{zz}$). The magnetic component dominates the dynamo, while the kinetic part remains subdominant throughout the evolution, as shown in \Fig{fig:BxdtBx_z_xyaver}(d).
	We now examine the physical origin of this mechanism, which operates through a negative $\eta_{yx}$, primarily its magnetic component $\eta_{yx}^{\text{mag}}$, in $\mathcal{\bar E}_y$. Specifically, we analyze how the term $\bar M_{zz} \partial_z \bar B_y$ contributes to $\mathcal{\bar E}_y$, thereby sustaining the mean field $\bar B_x$.


	\begin{figure}
		\includegraphics[width=\columnwidth]{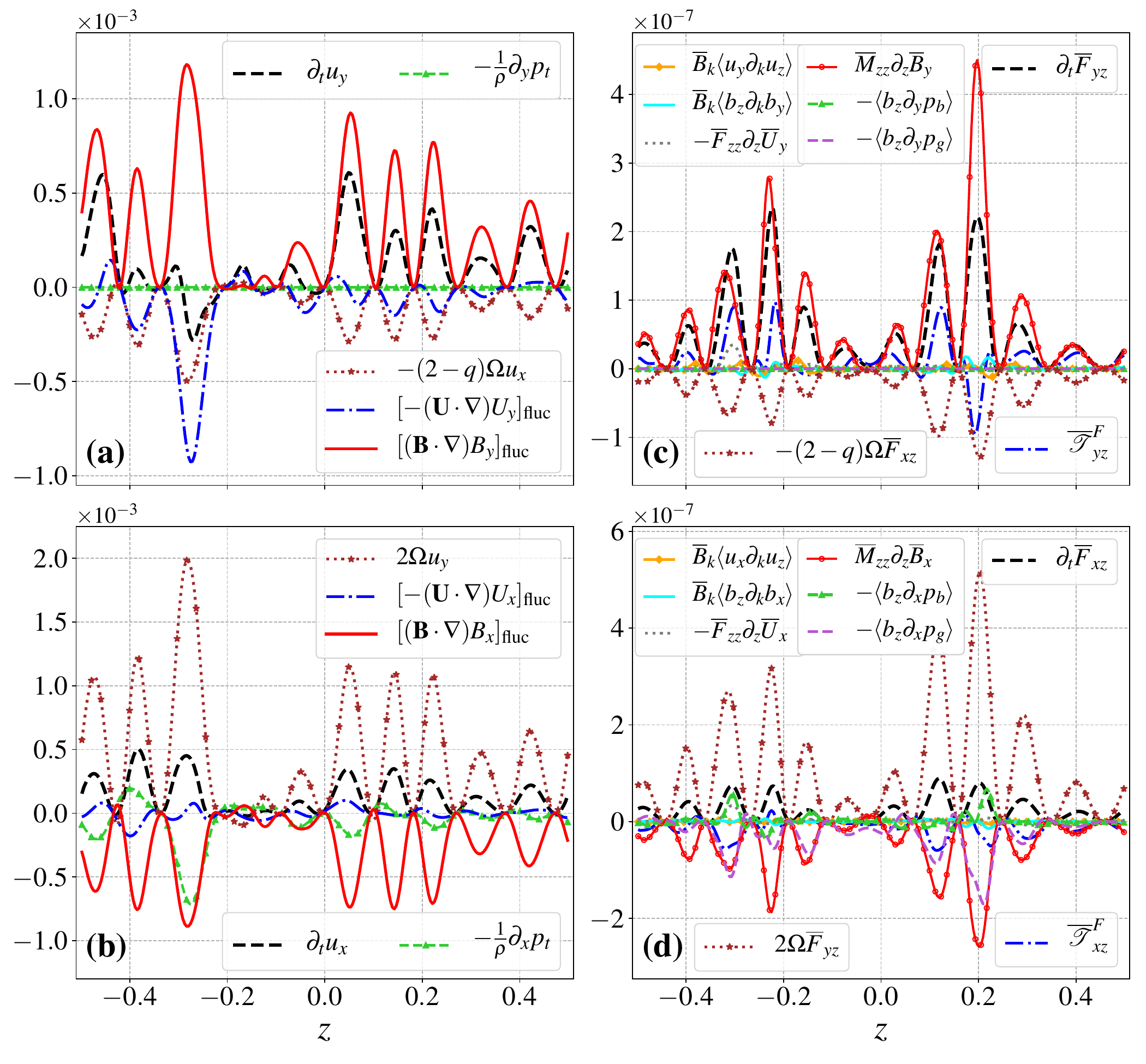}  
		\caption{
			Spatial variation of individual terms in the evolution equations for fluctuating velocity fields and mean Faraday stress during the MRI growth phase, evaluated at $t/T_{\text{orb}} =3.34$. Panels (a, b) show terms for $u_y$ and $u_x$ (Eq.~\ref{eq:fluctuatingU}); panels (c, d) for $\bar F_{yz}$ and $\bar F_{xz}$ (Eq.~\ref{eq:Fij_exact}). 
			To ensure interpretability, we multiply $u_i$ and $\bar F_{ij}$ on both sides of their respective equations so that source terms remain positive (not shown in legends).
		}
		\label{fig:ui_Fij_x-0p25_y0_T3p34}
	\end{figure}
	In the presence of both mean ($\bar J_x = - \partial_z \bar B_y$) and fluctuating ($j_x = - \partial_z b_y$) currents, magnetic fluctuations $b_z$ interact with these field gradients to drive velocity fluctuations $u_y$ via the magnetic tension term $[(\bm B \cdot \nabla) B_y]_\text{fluc}$ in the fluctuating Lorentz force, particularly
	\begin{equation}
		u_y \sim \frac{\tau_y}{\mu_0 \rho} (b_z \partial_z \bar B_y + b_z \partial_z b_y) .
	\end{equation}
	The Coriolis force subsequently converts $u_y$ into $u_x$: 
	\begin{equation}
		u_x \sim \tau_x 2\Omega u_y \sim 2 \Omega \tau_x \frac{\tau_y}{\mu_0 \rho} (b_z \partial_z \bar B_y + b_z \partial_z b_y),
	\end{equation}
	where, $\tau_i$ is the turbulent correlation time. 
	\Fig{fig:ui_Fij_x-0p25_y0_T3p34} illustrates the contribution of individual terms in the $u_i$ evolution equations (see Appendix~\ref{sec:appendix_EMF}). 
	To track the energy 
	we multiply $u_i$ on both sides of the $\partial_t u_i$ equations. 
	The magnetic tension term of Lorentz force fluctuations drives $u_y$ (\Fig{fig:ui_Fij_x-0p25_y0_T3p34}a), which is then converted into $u_x$ by the Coriolis force (\Fig{fig:ui_Fij_x-0p25_y0_T3p34}b). Contributions from pressure fluctuations are negligible for $u_y$, and remain small (mostly dissipative) in the $u_x$ evolution.

	The resulting $u_x$ correlates with its source $b_z$, producing an EMF $\mathcal{\bar E}_y = \left< u_z b_x \right> - \left< u_x b_z \right>$, that is proportional to and has the same sign as $\bar J_x$:
	\begin{align}
		\mathcal{\bar E}_y &\sim -\left< u_x b_z \right> \sim - \tau_x 2\Omega \left< u_y b_z \right> \nonumber\\
		&\sim -2 \Omega \tau_x (\bar{M}_{zz} \partial_z \bar B_y + \left< b_z b_z \partial_z b_y \right>)  \nonumber\\
		& \sim 2 \Omega \tau_x (\bar{M}_{zz} \bar J_x - \left< b_z b_z \partial_z b_y \right>),
	\end{align}
	where the second term on the RHS is a third-order correlator.  Through its correlation with positive $\bar{M}_{zz}$,  $\bar J_x$  generates $\mathcal{\bar E}_y$, responsible for the growth of $\bar B_x$. This process underpins the origin of the negative $\eta_{yx}^{\text{mag}}$, through which the large-scale dynamo mechanism---rotation-shear-current effect---operates.
	Meanwhile,  $u_y$ correlates with $b_z$ to produce   $\mathcal{\bar E}_x = \left< u_y b_z \right> - \left< u_z b_y \right>$, which is proportional to but opposite in sign to $\bar J_x$:
	\begin{align}
		\mathcal{\bar E}_x \sim \left< u_y b_z \right> &\sim \tau_y (\bar{M}_{zz} \partial_z \bar B_y + \left< b_z b_z \partial_z b_y \right>) \nonumber\\
		& \sim \tau_y (-\bar{M}_{zz} \bar J_x + \left< b_z b_z \partial_z b_y \right>).
	\end{align}
	Thus, $\mathcal{\bar E}_x$ acts as a turbulent dissipation mechanism for the mean field $\bar B_y$, counteracting its growth.
	Figs.~\ref{fig:ui_Fij_x-0p25_y0_T3p34}(c) and (d) illustrate the contributions of individual terms in the evolution equations (see Appendix~\ref{sec:appendix_EMF}) for $\bar{F}_{yz} = \left< u_y b_z \right>$ and $\bar{F}_{xz} = \left< u_x b_z \right>$, evaluated during the MRI exponential growth phase at $t/T_{\text{orb}} =3.34$. The dominant source term for $\bar{F}_{yz}$ is $\bar{M}_{zz} \partial_z \bar B_y$ (\Fig{fig:ui_Fij_x-0p25_y0_T3p34}c), with the Coriolis force converting $\bar{F}_{yz}$ into $\bar{F}_{xz}$ (\Fig{fig:ui_Fij_x-0p25_y0_T3p34}d). In both cases, terms proportional to $\bar B_i$ are negligible. While azimuthal gas and magnetic pressure gradients do not contribute to $\bar{F}_{yz}$, a weak radial pressure gradient affects $\bar{F}_{xz}$.
	

	To support the physical interpretation of the rotation-shear-current effect, we present visualizations of fluctuating velocity fields from DNS.
	\begin{figure}
		\includegraphics[width=1.0\columnwidth]{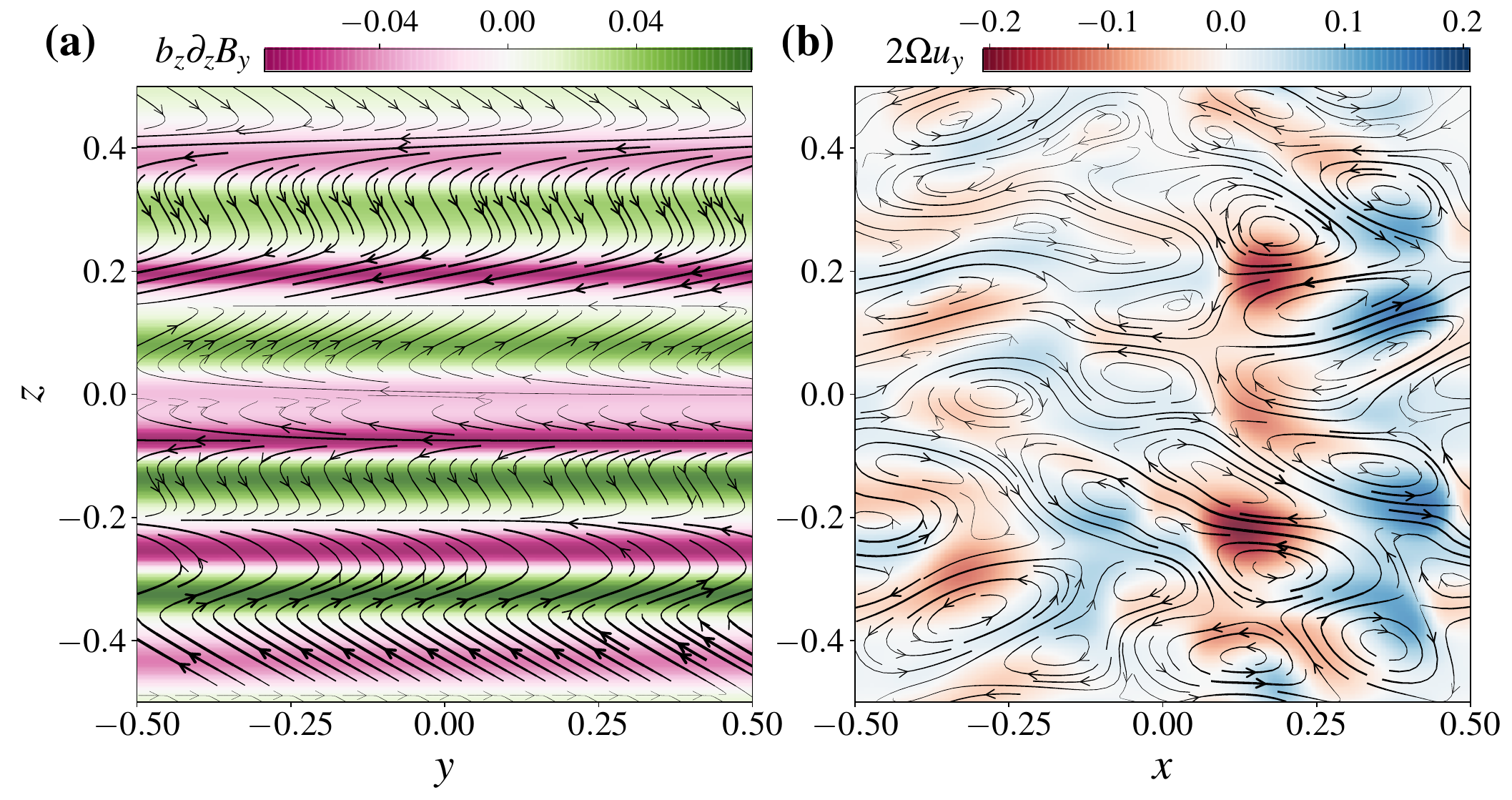}  
		\caption{	 
			Streamlines of in-plane fluctuating velocity fields during the exponential growth phase of MRI at $t/T_{\text{orb}} =3.34$.
			(a) Streamlines of ($u_y$, $u_z$) in the y-z plane at $x=0.25$, overlaid on a colormap of the magnetic tension term $b_z \partial_z B_y$, which drives $u_y$.
			(b) Streamlines of ($u_x$, $u_z$) in the x-z plane at $y=0.25$, with a colormap of the Coriolis term, which converts $u_y$ into $u_x$.
		}
		\label{fig:uy_ux_streamline}
	\end{figure}
	Figure~\ref{fig:uy_ux_streamline} shows streamlines of fluctuating velocities, illustrating how magnetic tension drives $u_y$, which is subsequently converted into $u_x$ by the Coriolis force. In panel (a), ($u_y$, $u_z$) streamlines in the y-z plane at $x=0.25$ are overlaid on a colormap of the magnetic tension term $b_z \partial_z B_y$ from the $u_y$ equation. The streamlines, along the y-direction, closely follow regions of strong magnetic tension, confirming its role as the primary driver of $u_y$. Panel (b) shows ($u_x$, $u_z$) streamlines in the x-z plane at $y=0.25$, with a colormap of the Coriolis term from the $u_x$ equation. The alignment of streamlines with regions of strong Coriolis acceleration demonstrates the efficient conversion of $u_y$ into $u_x$.

	\begin{figure*}
		\includegraphics[width=1.9\columnwidth]{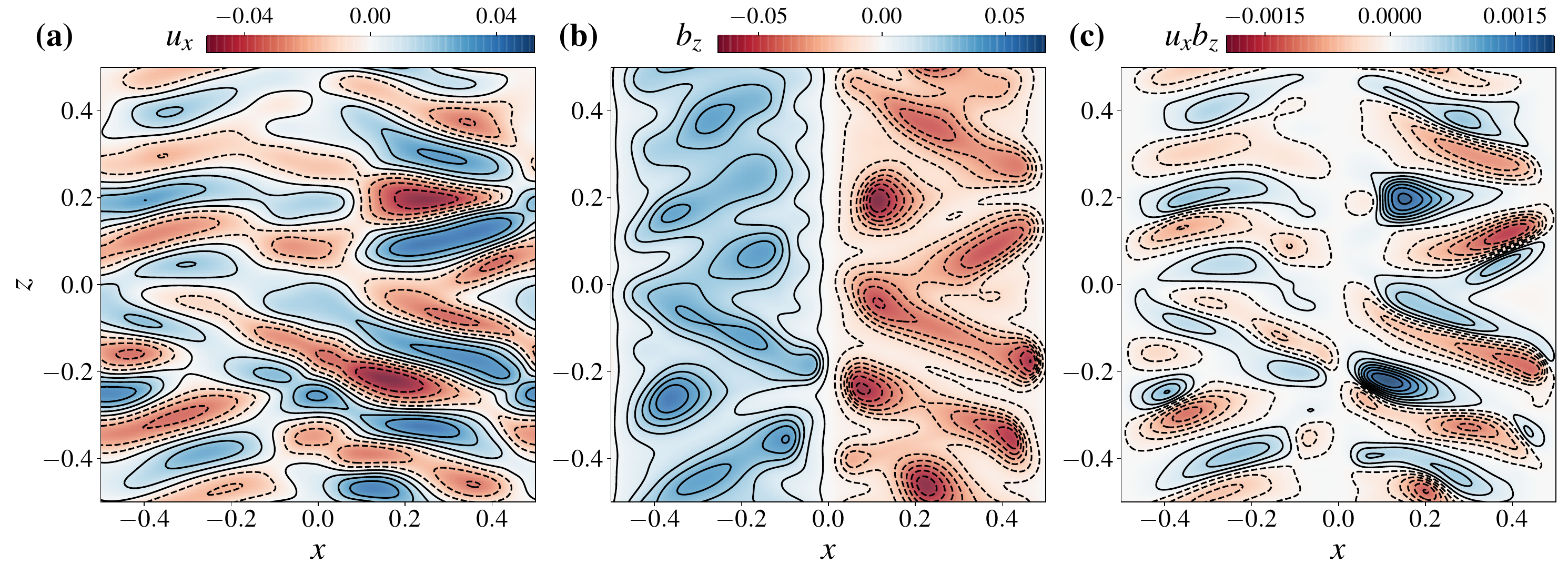}  
		\caption{	 
			Colormaps with contours of fluctuating fields in the x-z plane during the exponential growth phase of MRI, at $t/T_{\text{orb}} =3.34$.
			(a) Fluctuating velocity $u_x$, (b) fluctuating magnetic field $b_z$, and (c) their product $u_x b_z$, with solid (dashed) contours denoting positive (negative) values.
			While fluctuations $u_x$ and $b_z$ individually average out when integrated over $x$, their correlator $u_x b_z$ remains finite, indicating a nonzero mean stress.
		}
		\label{fig:ux_bz_contour}
	\end{figure*}
	To connect fluctuating fields to the mean EMF, \Fig{fig:ux_bz_contour} shows the relevant fluctuating fields and their correlator. Panel (a) displays the fluctuating velocity $u_x$; panel (b), the fluctuating magnetic field $b_z$, and panel (c), their product $u_x b_z$, which contributes to the EMF component $\mathcal{\bar E}_y$. Although $u_x$ and $b_z$ individually average out when integrated over $x$---being nearly independent of $y$ during the exponential growth phase---their product remains finite and structured, revealing a coherent correlation that sustains a nonzero EMF, $\mathcal{\bar E}_y (z) \sim - \left< u_x b_z \right>$. This EMF is proportional to, and shares the sign of $\bar J_x$, reinforcing the generation of the mean field $\bar B_x$ discussed earlier.


	\begin{figure}
		\includegraphics[width=\columnwidth]{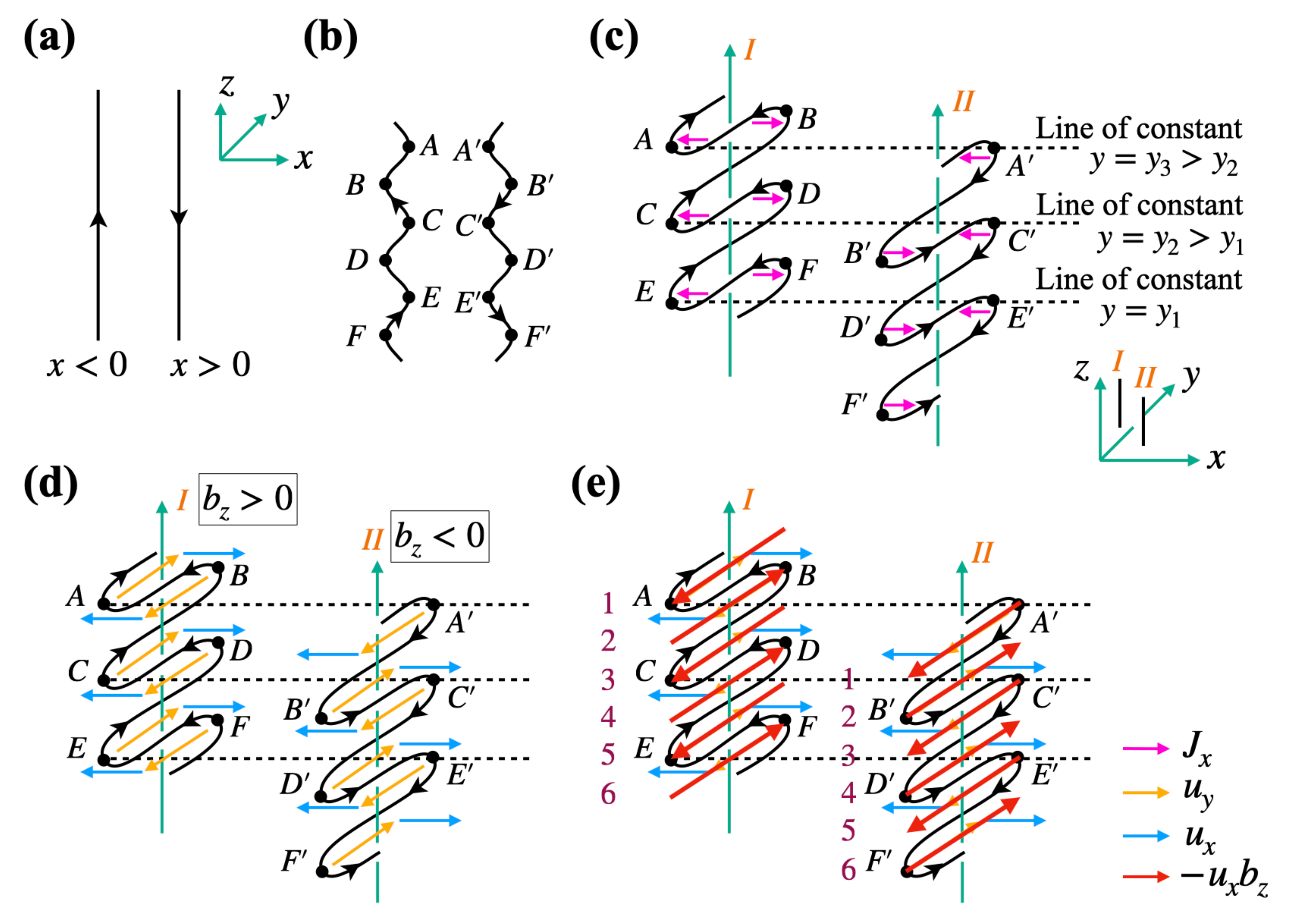}  
		\caption{Schematic illustration of the rotation-shear-current effect. (a) Two initially vertical magnetic field sectors with opposite polarity. (b) An $x$-dependent perturbation with a phase shift is introduced. (c) Background shear $(\partial U^0_y / \partial x < 0)$ stretches the field lines and shifts the sectors relative to each other. The resulting field line bending produces radial currents $J_x \sim - \partial_z B_y$. (d) Magnetic tension from field curvature induces $u_y$ fluctuations, which are converted into $u_x$ via the Coriolis force. (e) The resulting vectors of $- u_x b_z$  are shown, with $z$-layers (1 to 6) indicating matched heights across stacks. The EMF component $\mathcal{E}_y \sim - u_x b_z$ aligns within each $x$-$y$ plane at fixed $z$, but reverses sign between adjacent $z$-layers. After $xy$-averaging, $\mathcal{\bar E}_y (z)$ exhibits an alternating pattern: $\mathcal{\bar E}_y < 0$ at layer 1, $\mathcal{\bar E}_y > 0$ at layer 2, $\mathcal{\bar E}_y < 0$ at layer 3, and so on. The induced mean-field satisfies $\partial_t \bar B_x = - \partial_z \mathcal{\bar E}_y$, generating a coherent $\bar B_x (z)$ that reverses in $z$, consistent with the fastest-growing MRI mode.
		}
		\label{fig:schematic}
	\end{figure}
	Finally, the mechanism underlying the rotation-shear-current effect is illustrated schematically in \Fig{fig:schematic}. Initially (panel a), two oppositely directed vertical magnetic field sectors are placed side by side (see Supplemental Material \ref{sec:method} for simulation setup). A small perturbation with a phase shift in $x$ is introduced (panel b). Background shear $(\partial U^0_y / \partial x < 0)$ stretches the field lines and displaces the two sectors relative to each other (panel c). The bending of field lines induces radial currents $J_x \sim -\partial_z B_y$. The associated field curvature produces a magnetic tension force that drives velocity fluctuations $u_y$, which are subsequently redirected into $u_x$ by the Coriolis force (panel d). These fluctuations produce an EMF component $\mathcal{E}_y \sim - u_x b_z$ (panel e). Notably, $\mathcal{\bar E}_y$ aligns coherently within each $x$-$y$ plane but alternates in sign across $z$-layers. This results in a layerwise alternating EMF profile, $\mathcal{\bar E}_y (z)$, which drives the growth of a large-scale magnetic field $\bar B_x (z)$ (following Eq.~\ref{eq:meanBx_xy_2}) that reverses in $z$, consistent with the fastest-growing MRI mode.


	In summary,  we have demonstrated that, unlike traditional $\alpha$-$\Omega$ dynamos that rely on helicity and stratification, large-scale dynamos in unstratified, zero-net-flux MRI-unstable disks arise through a fundamentally different mechanism---the rotation-shear-current effect.
	This dynamo operates through a negative off-diagonal turbulent diffusivity, $\eta_{yx}$, which amplifies large-scale radial magnetic fields. Horizontal planar averaging reveals that $\mathcal{\bar E}_y$ drives the generation of $\bar B_x$, which is then stretched into $\bar B_y$ by shear via the $\Omega$-effect, completing the dynamo cycle.
	The dominant source of $\mathcal{\bar E}_y$ is the magnetic contribution to turbulent diffusivity, $\eta_{yx}^{\text{mag}}$, arising from the interplay between fluctuating Lorentz and Coriolis forces. Specifically, magnetic tension $b_z \partial_z \bar B_y$ drives $u_y$ via interactions between vertical magnetic fluctuations $b_z$ and the mean current $\bar J_x$. The Coriolis force then converts $u_y$ into $u_x$, which in turn correlates with $b_z$ to produce an EMF $\mathcal{\bar E}_y \sim - \left< u_x b_z \right>$. This EMF is proportional to and has the same sign as $\bar J_x$, reinforcing the growth of $\bar B_x$ and sustaining the dynamo. After the initial exponential growth, third-order correlators arising from Lorentz force fluctuations counteract the $\eta_{yx}$-driven amplification, saturating the dynamo in fully developed MRI turbulence.

	Crucially, this work introduces a fundamentally new approach by constructing the mean EMF self-consistently from the evolution equations of Faraday tensors, enabling explicit identification of turbulent transport coefficients without any closure approximations or fitting procedures. By directly analyzing the evolution of fluctuating fields that constitute the EMF correlators, we uncover the complete production chain underlying large-scale magnetic field generation in MRI turbulence. This framework establishes a physically grounded method for diagnosing dynamo mechanisms in rotating shear flows as well as in global models with curvature effects~\cite{ebrahimi2022nonlocal,ebrahimi2025generalized}.

	\vspace{10pt}
	\noindent {\bf Acknowledgments}:
	
	The simulations were performed on the International Centre for Theoretical Sciences (ICTS) HPC cluster \textit{Contra}. T.M. and P.B. acknowledge support of the Department of Atomic Energy, Government of India, under project no. RTI4001. E.B. acknowledges support from National Science Foundation (NSF) Physics Frontier Center grant PHY-2020249. F.E. acknowledges support from NSF under Award No. 2308839.

	
	\clearpage
	
	\appendix
	
	\begin{widetext}
		\section{The Electromotive Force} \label{sec:appendix_EMF}
		
		In this section, we provide a comprehensive derivation of the electromotive force (EMF). Following standard mean-field theory, we apply a Reynolds decomposition to the dynamical flow variables, expressing each as the sum of a mean component (denoted by over-bars) and a fluctuating component (represented by lowercase letters): $B_i = \bar B_i + b_i$. This decomposition satisfies Reynolds averaging rules, i.e., $\bar b_i = 0, \ \bar{\bar B}_i = \bar B_i$.
		By subtracting the mean (in this case, the xy-averaged) equation from the total, we obtain the evolution equations for the fluctuating velocity and magnetic fields:
		\begin{eqnarray}
			\mathcal{D}_t u_i &=& q\Omega u_x \hat{y} - 2\epsilon_{ijk}\Omega_j u_k - [(\bec U \cdot \nabla)U_i]_{\text{fluc}} + \frac{1}{\mu_0 \rho} [(\bec B \cdot \nabla)B_i]_{\text{fluc}} - \frac{1}{\rho}\partial_i p_t + + \nu\partial_{jj} u_i \nonumber\\
			&=& q\Omega u_x \hat{y} - 2\epsilon_{ijk}\Omega_j u_k -\bar U_j\partial_j u_i -u_j\partial_j \bar U_i  
			+ \frac{1}{\mu_0 \rho}(\bar B_j\partial_j b_i + b_j\partial_j\bar B_i) 
			+ \frac{1}{\rho}\partial_j(M_{ij}-R_{ij}-\bar M_{ij}+\bar R_{ij}) \nonumber\\
			&& - \frac{1}{\rho}\partial_i p_t + \nu\partial_{jj} u_i \;,
			\label{eq:fluctuatingU} \\
			\mathcal{D}_t b_i  &=& - q\Omega b_x \hat{y} + \bar B_j\partial_j u_i + b_j\partial_j\bar U_i -\bar U_j\partial_j b_i - u_j\partial_j\bar B_i 
			+ \partial_j (F_{ij}-F_{ji}-\bar F_{ij}+\bar F_{ji}) +\eta\partial_{jj} b_i \;.
			\label{eq:fluctuatingB}
		\end{eqnarray}
		Here, $\mathcal{D}_t =  \partial_t - q\Omega x \partial_y$ includes advection by the background shear. $M_{ij}=b_i b_j / \mu_0$, $R_{ij}=\rho u_i u_j$, and $F_{ij}=u_i b_j$ represent the Maxwell, Reynolds, and Faraday tensors, respectively.
		Using \Eqs{eq:fluctuatingU}{eq:fluctuatingB}, and by applying Reynolds averaging rules, we obtain the evolution equation for the mean Faraday tensors:
		\begin{multline}
			\mathcal{D}_t \bar F_{ij} + \bar U_k\partial_k \bar F_{ij} - \bar F_{ik}\partial_k\bar U_j + \bar F_{kj}\partial_k\bar U_i + 2\epsilon_{ikl}\Omega_k\bar F_{lj} 
			+ \bar S_{ij}^F - \frac{1}{\rho}( \bar M_{jk}\partial_k\bar B_i-\bar R_{ik}\partial_k\bar B_j) \\ = -\frac{1}{\rho}\langle b_j\partial_i p_t \rangle + \bar B_k \langle u_i\partial_k u_j \rangle  
			+ \frac{\bar B_k}{\mu_0 \rho}  \langle b_j\partial_k b_i \rangle + \mathcal{\bar T}^F_{ij} 
			+ \eta \langle u_i\partial_{kk}b_j \rangle + \nu \langle b_j\partial_{kk}u_i\rangle. 
			\label{eq:Fij_exact} 
		\end{multline}
		Here, $\bar S_{ij}^F =  - \bar F_{ik}\partial_k\bar U^0_j + \bar F_{kj}\partial_k\bar U^0_i $ describes how the Faraday tensors are `stretched' by the background shear flow, ${\bec U^0}$, while
		$\mathcal{\bar T}^F_{ij} = \langle u_i b_k \partial_k u_j + b_j b_k \partial_k b_i - u_k \partial_k F_{ij} \rangle $ represents the nonlinear third-order correlator.
		To construct the EMF, we utilize interaction terms from the Coriolis force and background shear in the evolution equations for the Faraday tensors. Our primary focus is on the azimuthal component of the EMF, given by: $\mathcal{\bar E}_y = \bar F_{zx} - \bar F_{xz}$. 
		The evolution equations for the relevant Faraday tensor components are:
		\begin{align}	
			\mathcal{D}_t \bar F_{yz} =& -(2-q)\Omega \bar F_{xz} - \frac{1}{\rho}\langle b_z\partial_y p_t \rangle
			+ \left(\bar F_{yk}\partial_k\bar U_z - \bar F_{kz}\partial_k\bar U_y \right)
			+ \bar B_k \left[ \langle u_y\partial_k u_z \rangle  + \frac{\langle b_z\partial_k b_y \rangle}{\mu_0 \rho} \right] \nonumber\\
			& + \frac{1}{\rho}\left( \bar M_{zk}\partial_k\bar B_y-\bar R_{yk}\partial_k\bar B_z \right) + \mathcal{\bar T}^F_{yz}, 
			\label{eq:F_yz}  \\
			\mathcal{D}_t \bar F_{zy} =& - q\Omega \bar F_{zx} - \frac{1}{\rho}\langle b_y\partial_z p_t \rangle
			+ \left(\bar F_{zk}\partial_k\bar U_y - \bar F_{ky}\partial_k\bar U_z \right)
			+ \bar B_k \left[ \langle u_z\partial_k u_y \rangle  + \frac{\langle b_y\partial_k b_z \rangle}{\mu_0 \rho} \right] \nonumber\\
			& + \frac{1}{\rho}\left( \bar M_{yk}\partial_k\bar B_z-\bar R_{zk}\partial_k\bar B_y \right) + \mathcal{\bar T}^F_{zy} . 
			\label{eq:F_zy}
		\end{align}
		To derive $\mathcal{\bar E}_y$, we multiply \Eq{eq:F_yz} by $q$ and \Eq{eq:F_zy} by $(q-2)$, and subsequently combine them. After algebraic simplifications, we obtain:
		\begin{align}
			q \mathcal{D}_t \bar F_{yz} + (q-2) \mathcal{D}_t \bar F_{zy} = & \ q(2-q) \Omega (\bar F_{zx} - \bar F_{xz}) 
			+ \left \{ q\bar F_{yk} +(2-q)\bar F_{ky} \right \} \partial_k \bar U_z     
			- \left\{ q\bar F_{kz} +(2-q)\bar F_{zk} \right \} \partial_k \bar U_y         \nonumber\\
			& + \frac{1}{\rho}\left \{q\bar M_{zk} +(2-q)\bar R_{zk} \right\} \partial_k \bar B_y
			- \frac{1}{\rho}\left\{q\bar R_{yk} +(2-q)\bar M_{yk} \right\}\partial_k \bar B_z  \nonumber\\
			& + \bar B_k \left\{ q \left(\langle u_y \partial_k u_z \rangle + \frac{\langle b_z \partial_k b_y \rangle}{\mu_0 \rho} \right) - (2-q) \left(\langle u_z \partial_k u_y \rangle + \frac{\langle b_y \partial_k b_z \rangle}{\mu_0 \rho} \right) \right\}  \nonumber\\
			& - \frac{1}{\rho} \Big\{ q \langle b_z\partial_y p_t \rangle - (2-q) \langle b_y\partial_z p_t \rangle \Big\} + \mathcal{\bar T}_{y} ,
		\end{align}
		where, $\mathcal{\bar T}_{y} = q \mathcal{\bar T}^F_{yz} - (2-q) \mathcal{\bar T}^F_{zy}$ represents the contribution from third-order correlators. By further algebraic manipulation, we arrive at the final expression for $\mathcal{\bar E}_y = (\bar F_{zx} - \bar F_{xz})$:
		\begin{align}
			\mathcal{\bar E}_y = \frac{-1}{q(2-q)\Omega} \Bigg[
			& \left\{ - q \mathcal{D}_t \bar F_{yz} + (2-q) \mathcal{D}_t \bar F_{zy} \right\}
			+ \left \{ q\bar F_{yk} +(2-q)\bar F_{ky} \right \} \partial_k \bar U_z     
			- \left\{ q\bar F_{kz} +(2-q)\bar F_{zk} \right \} \partial_k \bar U_y         \nonumber\\
			& + \frac{1}{\rho}\left \{q\bar M_{zk} +(2-q)\bar R_{zk} \right\} \partial_k \bar B_y
			- \frac{1}{\rho}\left\{q\bar R_{yk} +(2-q)\bar M_{yk} \right\}\partial_k \bar B_z  \nonumber\\
			& + \bar B_k \left\{ q \left(\langle u_y \partial_k u_z \rangle + \frac{\langle b_z \partial_k b_y \rangle}{\mu_0 \rho} \right) - (2-q) \left(\langle u_z \partial_k u_y \rangle + \frac{\langle b_y \partial_k b_z \rangle}{\mu_0 \rho} \right) \right\}  \nonumber\\
			& - \frac{1}{\rho} \Big\{ q \langle b_z\partial_y p_t \rangle - (2-q) \langle b_y\partial_z p_t \rangle 
			+ \mathcal{\bar T}_{y} \Bigg] .
		\end{align}
		
	\end{widetext}
	
	
	\bibliography{mri_dynamo}

\begin{thebibliography}{38}%
\makeatletter
\providecommand \@ifxundefined [1]{%
 \@ifx{#1\undefined}
}%
\providecommand \@ifnum [1]{%
 \ifnum #1\expandafter \@firstoftwo
 \else \expandafter \@secondoftwo
 \fi
}%
\providecommand \@ifx [1]{%
 \ifx #1\expandafter \@firstoftwo
 \else \expandafter \@secondoftwo
 \fi
}%
\providecommand \natexlab [1]{#1}%
\providecommand \enquote  [1]{``#1''}%
\providecommand \bibnamefont  [1]{#1}%
\providecommand \bibfnamefont [1]{#1}%
\providecommand \citenamefont [1]{#1}%
\providecommand \href@noop [0]{\@secondoftwo}%
\providecommand \href [0]{\begingroup \@sanitize@url \@href}%
\providecommand \@href[1]{\@@startlink{#1}\@@href}%
\providecommand \@@href[1]{\endgroup#1\@@endlink}%
\providecommand \@sanitize@url [0]{\catcode `\\12\catcode `\$12\catcode
  `\&12\catcode `\#12\catcode `\^12\catcode `\_12\catcode `\%12\relax}%
\providecommand \@@startlink[1]{}%
\providecommand \@@endlink[0]{}%
\providecommand \url  [0]{\begingroup\@sanitize@url \@url }%
\providecommand \@url [1]{\endgroup\@href {#1}{\urlprefix }}%
\providecommand \urlprefix  [0]{URL }%
\providecommand \Eprint [0]{\href }%
\providecommand \doibase [0]{https://doi.org/}%
\providecommand \selectlanguage [0]{\@gobble}%
\providecommand \bibinfo  [0]{\@secondoftwo}%
\providecommand \bibfield  [0]{\@secondoftwo}%
\providecommand \translation [1]{[#1]}%
\providecommand \BibitemOpen [0]{}%
\providecommand \bibitemStop [0]{}%
\providecommand \bibitemNoStop [0]{.\EOS\space}%
\providecommand \EOS [0]{\spacefactor3000\relax}%
\providecommand \BibitemShut  [1]{\csname bibitem#1\endcsname}%
\let\auto@bib@innerbib\@empty
\bibitem [{\citenamefont {{Balbus}}\ and\ \citenamefont
  {{Hawley}}(1998)}]{1998RvMP...70....1B}%
  \BibitemOpen
  \bibfield  {author} {\bibinfo {author} {\bibfnamefont {S.~A.}\ \bibnamefont
  {{Balbus}}}\ and\ \bibinfo {author} {\bibfnamefont {J.~F.}\ \bibnamefont
  {{Hawley}}},\ }\bibfield  {title} {\bibinfo {title} {{Instability,
  turbulence, and enhanced transport in accretion disks}},\ }\href
  {https://doi.org/10.1103/RevModPhys.70.1} {\bibfield  {journal} {\bibinfo
  {journal} {Reviews of Modern Physics}\ }\textbf {\bibinfo {volume} {70}},\
  \bibinfo {pages} {1} (\bibinfo {year} {1998})}\BibitemShut {NoStop}%
\bibitem [{\citenamefont {{Bhat}}\ \emph {et~al.}(2017)\citenamefont {{Bhat}},
  \citenamefont {{Ebrahimi}}, \citenamefont {{Blackman}},\ and\ \citenamefont
  {{Subramanian}}}]{2017MNRAS.472.2569B}%
  \BibitemOpen
  \bibfield  {author} {\bibinfo {author} {\bibfnamefont {P.}~\bibnamefont
  {{Bhat}}}, \bibinfo {author} {\bibfnamefont {F.}~\bibnamefont {{Ebrahimi}}},
  \bibinfo {author} {\bibfnamefont {E.~G.}\ \bibnamefont {{Blackman}}},\ and\
  \bibinfo {author} {\bibfnamefont {K.}~\bibnamefont {{Subramanian}}},\
  }\bibfield  {title} {\bibinfo {title} {{Evolution of the magnetorotational
  instability on initially tangled magnetic fields}},\ }\href
  {https://doi.org/10.1093/mnras/stx1989} {\bibfield  {journal} {\bibinfo
  {journal} {\mnras}\ }\textbf {\bibinfo {volume} {472}},\ \bibinfo {pages}
  {2569} (\bibinfo {year} {2017})}\BibitemShut {NoStop}%
\bibitem [{\citenamefont {{Brandenburg}}\ \emph {et~al.}(1995)\citenamefont
  {{Brandenburg}}, \citenamefont {{Nordlund}}, \citenamefont {{Stein}},\ and\
  \citenamefont {{Torkelsson}}}]{1995ApJ...446..741B}%
  \BibitemOpen
  \bibfield  {author} {\bibinfo {author} {\bibfnamefont {A.}~\bibnamefont
  {{Brandenburg}}}, \bibinfo {author} {\bibfnamefont {A.}~\bibnamefont
  {{Nordlund}}}, \bibinfo {author} {\bibfnamefont {R.~F.}\ \bibnamefont
  {{Stein}}},\ and\ \bibinfo {author} {\bibfnamefont {U.}~\bibnamefont
  {{Torkelsson}}},\ }\bibfield  {title} {\bibinfo {title} {{Dynamo-generated
  Turbulence and Large-Scale Magnetic Fields in a Keplerian Shear Flow}},\
  }\href {https://doi.org/10.1086/175831} {\bibfield  {journal} {\bibinfo
  {journal} {\apj}\ }\textbf {\bibinfo {volume} {446}},\ \bibinfo {pages} {741}
  (\bibinfo {year} {1995})}\BibitemShut {NoStop}%
\bibitem [{\citenamefont {{Hawley}}\ \emph {et~al.}(1996)\citenamefont
  {{Hawley}}, \citenamefont {{Gammie}},\ and\ \citenamefont
  {{Balbus}}}]{1996ApJ...464..690H}%
  \BibitemOpen
  \bibfield  {author} {\bibinfo {author} {\bibfnamefont {J.~F.}\ \bibnamefont
  {{Hawley}}}, \bibinfo {author} {\bibfnamefont {C.~F.}\ \bibnamefont
  {{Gammie}}},\ and\ \bibinfo {author} {\bibfnamefont {S.~A.}\ \bibnamefont
  {{Balbus}}},\ }\bibfield  {title} {\bibinfo {title} {{Local Three-dimensional
  Simulations of an Accretion Disk Hydromagnetic Dynamo}},\ }\href
  {https://doi.org/10.1086/177356} {\bibfield  {journal} {\bibinfo  {journal}
  {\apj}\ }\textbf {\bibinfo {volume} {464}},\ \bibinfo {pages} {690} (\bibinfo
  {year} {1996})}\BibitemShut {NoStop}%
\bibitem [{\citenamefont {{Lesur}}\ and\ \citenamefont
  {{Ogilvie}}(2008)}]{2008A&A...488..451L}%
  \BibitemOpen
  \bibfield  {author} {\bibinfo {author} {\bibfnamefont {G.}~\bibnamefont
  {{Lesur}}}\ and\ \bibinfo {author} {\bibfnamefont {G.~I.}\ \bibnamefont
  {{Ogilvie}}},\ }\bibfield  {title} {\bibinfo {title} {{On self-sustained
  dynamo cycles in accretion discs}},\ }\href
  {https://doi.org/10.1051/0004-6361:200810152} {\bibfield  {journal} {\bibinfo
   {journal} {\aap}\ }\textbf {\bibinfo {volume} {488}},\ \bibinfo {pages}
  {451} (\bibinfo {year} {2008})}\BibitemShut {NoStop}%
\bibitem [{\citenamefont {{Gressel}}(2010)}]{2010MNRAS.405...41G}%
  \BibitemOpen
  \bibfield  {author} {\bibinfo {author} {\bibfnamefont {O.}~\bibnamefont
  {{Gressel}}},\ }\bibfield  {title} {\bibinfo {title} {{A mean-field approach
  to the propagation of field patterns in stratified magnetorotational
  turbulence}},\ }\href {https://doi.org/10.1111/j.1365-2966.2010.16440.x}
  {\bibfield  {journal} {\bibinfo  {journal} {\mnras}\ }\textbf {\bibinfo
  {volume} {405}},\ \bibinfo {pages} {41} (\bibinfo {year} {2010})}\BibitemShut
  {NoStop}%
\bibitem [{\citenamefont {{Bhat}}\ \emph {et~al.}(2016)\citenamefont {{Bhat}},
  \citenamefont {{Ebrahimi}},\ and\ \citenamefont
  {{Blackman}}}]{2016MNRAS.462..818B}%
  \BibitemOpen
  \bibfield  {author} {\bibinfo {author} {\bibfnamefont {P.}~\bibnamefont
  {{Bhat}}}, \bibinfo {author} {\bibfnamefont {F.}~\bibnamefont {{Ebrahimi}}},\
  and\ \bibinfo {author} {\bibfnamefont {E.~G.}\ \bibnamefont {{Blackman}}},\
  }\bibfield  {title} {\bibinfo {title} {{Large-scale dynamo action precedes
  turbulence in shearing box simulations of the magnetorotational
  instability}},\ }\href {https://doi.org/10.1093/mnras/stw1619} {\bibfield
  {journal} {\bibinfo  {journal} {\mnras}\ }\textbf {\bibinfo {volume} {462}},\
  \bibinfo {pages} {818} (\bibinfo {year} {2016})}\BibitemShut {NoStop}%
\bibitem [{\citenamefont {{Mondal}}\ and\ \citenamefont
  {{Bhat}}(2023)}]{2023PhRvE.108f5201M}%
  \BibitemOpen
  \bibfield  {author} {\bibinfo {author} {\bibfnamefont {T.}~\bibnamefont
  {{Mondal}}}\ and\ \bibinfo {author} {\bibfnamefont {P.}~\bibnamefont
  {{Bhat}}},\ }\bibfield  {title} {\bibinfo {title} {{Unified treatment of
  mean-field dynamo and angular-momentum transport in magnetorotational
  instability-driven turbulence}},\ }\href
  {https://doi.org/10.1103/PhysRevE.108.065201} {\bibfield  {journal} {\bibinfo
   {journal} {\pre}\ }\textbf {\bibinfo {volume} {108}},\ \bibinfo {eid}
  {065201} (\bibinfo {year} {2023})}\BibitemShut {NoStop}%
\bibitem [{\citenamefont {{Krause}}\ and\ \citenamefont
  {{Raedler}}(1980)}]{1980opp..bookR....K}%
  \BibitemOpen
  \bibfield  {author} {\bibinfo {author} {\bibfnamefont {F.}~\bibnamefont
  {{Krause}}}\ and\ \bibinfo {author} {\bibfnamefont {K.~H.}\ \bibnamefont
  {{Raedler}}},\ }\href@noop {} {\emph {\bibinfo {title} {{Mean-field
  magnetohydrodynamics and dynamo theory}}}}\ (\bibinfo  {publisher} {Pergamon
  Press, Oxford},\ \bibinfo {year} {1980})\BibitemShut {NoStop}%
\bibitem [{\citenamefont {{Brandenburg}}\ and\ \citenamefont
  {{Subramanian}}(2005)}]{2005PhR...417....1B}%
  \BibitemOpen
  \bibfield  {author} {\bibinfo {author} {\bibfnamefont {A.}~\bibnamefont
  {{Brandenburg}}}\ and\ \bibinfo {author} {\bibfnamefont {K.}~\bibnamefont
  {{Subramanian}}},\ }\bibfield  {title} {\bibinfo {title} {{Astrophysical
  magnetic fields and nonlinear dynamo theory}},\ }\href
  {https://doi.org/10.1016/j.physrep.2005.06.005} {\bibfield  {journal}
  {\bibinfo  {journal} {\physrep}\ }\textbf {\bibinfo {volume} {417}},\
  \bibinfo {pages} {1} (\bibinfo {year} {2005})}\BibitemShut {NoStop}%
\bibitem [{\citenamefont {{Rincon}}(2019)}]{2019JPlPh..85d2001R}%
  \BibitemOpen
  \bibfield  {author} {\bibinfo {author} {\bibfnamefont {F.}~\bibnamefont
  {{Rincon}}},\ }\bibfield  {title} {\bibinfo {title} {{Dynamo theories}},\
  }\href {https://doi.org/10.1017/S0022377819000539} {\bibfield  {journal}
  {\bibinfo  {journal} {Journal of Plasma Physics}\ }\textbf {\bibinfo {volume}
  {85}},\ \bibinfo {eid} {205850401} (\bibinfo {year} {2019})}\BibitemShut
  {NoStop}%
\bibitem [{\citenamefont {{Tobias}}(2021)}]{2021JFM...912P...1T}%
  \BibitemOpen
  \bibfield  {author} {\bibinfo {author} {\bibfnamefont {S.~M.}\ \bibnamefont
  {{Tobias}}},\ }\bibfield  {title} {\bibinfo {title} {{The turbulent
  dynamo}},\ }\href {https://doi.org/10.1017/jfm.2020.1055} {\bibfield
  {journal} {\bibinfo  {journal} {Journal of Fluid Mechanics}\ }\textbf
  {\bibinfo {volume} {912}},\ \bibinfo {eid} {P1} (\bibinfo {year}
  {2021})}\BibitemShut {NoStop}%
\bibitem [{\citenamefont {{Rincon}}\ \emph {et~al.}(2007)\citenamefont
  {{Rincon}}, \citenamefont {{Ogilvie}},\ and\ \citenamefont
  {{Proctor}}}]{2007PhRvL..98y4502R}%
  \BibitemOpen
  \bibfield  {author} {\bibinfo {author} {\bibfnamefont {F.}~\bibnamefont
  {{Rincon}}}, \bibinfo {author} {\bibfnamefont {G.~I.}\ \bibnamefont
  {{Ogilvie}}},\ and\ \bibinfo {author} {\bibfnamefont {M.~R.~E.}\ \bibnamefont
  {{Proctor}}},\ }\bibfield  {title} {\bibinfo {title} {{Self-Sustaining
  Nonlinear Dynamo Process in Keplerian Shear Flows}},\ }\href
  {https://doi.org/10.1103/PhysRevLett.98.254502} {\bibfield  {journal}
  {\bibinfo  {journal} {\prl}\ }\textbf {\bibinfo {volume} {98}},\ \bibinfo
  {eid} {254502} (\bibinfo {year} {2007})}\BibitemShut {NoStop}%
\bibitem [{\citenamefont {{Herault}}\ \emph {et~al.}(2011)\citenamefont
  {{Herault}}, \citenamefont {{Rincon}}, \citenamefont {{Cossu}}, \citenamefont
  {{Lesur}}, \citenamefont {{Ogilvie}},\ and\ \citenamefont
  {{Longaretti}}}]{2011PhRvE..84c6321H}%
  \BibitemOpen
  \bibfield  {author} {\bibinfo {author} {\bibfnamefont {J.}~\bibnamefont
  {{Herault}}}, \bibinfo {author} {\bibfnamefont {F.}~\bibnamefont {{Rincon}}},
  \bibinfo {author} {\bibfnamefont {C.}~\bibnamefont {{Cossu}}}, \bibinfo
  {author} {\bibfnamefont {G.}~\bibnamefont {{Lesur}}}, \bibinfo {author}
  {\bibfnamefont {G.~I.}\ \bibnamefont {{Ogilvie}}},\ and\ \bibinfo {author}
  {\bibfnamefont {P.~Y.}\ \bibnamefont {{Longaretti}}},\ }\bibfield  {title}
  {\bibinfo {title} {{Periodic magnetorotational dynamo action as a prototype
  of nonlinear magnetic-field generation in shear flows}},\ }\href
  {https://doi.org/10.1103/PhysRevE.84.036321} {\bibfield  {journal} {\bibinfo
  {journal} {\pre}\ }\textbf {\bibinfo {volume} {84}},\ \bibinfo {eid} {036321}
  (\bibinfo {year} {2011})}\BibitemShut {NoStop}%
\bibitem [{\citenamefont {{Riols}}\ \emph {et~al.}(2017)\citenamefont
  {{Riols}}, \citenamefont {{Rincon}}, \citenamefont {{Cossu}}, \citenamefont
  {{Lesur}}, \citenamefont {{Ogilvie}},\ and\ \citenamefont
  {{Longaretti}}}]{2017A&A...598A..87R}%
  \BibitemOpen
  \bibfield  {author} {\bibinfo {author} {\bibfnamefont {A.}~\bibnamefont
  {{Riols}}}, \bibinfo {author} {\bibfnamefont {F.}~\bibnamefont {{Rincon}}},
  \bibinfo {author} {\bibfnamefont {C.}~\bibnamefont {{Cossu}}}, \bibinfo
  {author} {\bibfnamefont {G.}~\bibnamefont {{Lesur}}}, \bibinfo {author}
  {\bibfnamefont {G.~I.}\ \bibnamefont {{Ogilvie}}},\ and\ \bibinfo {author}
  {\bibfnamefont {P.~Y.}\ \bibnamefont {{Longaretti}}},\ }\bibfield  {title}
  {\bibinfo {title} {{Magnetorotational dynamo chimeras. The missing link to
  turbulent accretion disk dynamo models?}},\ }\href
  {https://doi.org/10.1051/0004-6361/201629285} {\bibfield  {journal} {\bibinfo
   {journal} {\aap}\ }\textbf {\bibinfo {volume} {598}},\ \bibinfo {eid} {A87}
  (\bibinfo {year} {2017})}\BibitemShut {NoStop}%
\bibitem [{\citenamefont {{Mamatsashvili}}\ \emph {et~al.}(2020)\citenamefont
  {{Mamatsashvili}}, \citenamefont {{Chagelishvili}}, \citenamefont {{Pessah}},
  \citenamefont {{Stefani}},\ and\ \citenamefont
  {{Bodo}}}]{2020ApJ...904...47M}%
  \BibitemOpen
  \bibfield  {author} {\bibinfo {author} {\bibfnamefont {G.}~\bibnamefont
  {{Mamatsashvili}}}, \bibinfo {author} {\bibfnamefont {G.}~\bibnamefont
  {{Chagelishvili}}}, \bibinfo {author} {\bibfnamefont {M.~E.}\ \bibnamefont
  {{Pessah}}}, \bibinfo {author} {\bibfnamefont {F.}~\bibnamefont
  {{Stefani}}},\ and\ \bibinfo {author} {\bibfnamefont {G.}~\bibnamefont
  {{Bodo}}},\ }\bibfield  {title} {\bibinfo {title} {{Zero Net Flux MRI
  Turbulence in Disks: Sustenance Scheme and Magnetic Prandtl Number
  Dependence}},\ }\href {https://doi.org/10.3847/1538-4357/abbd42} {\bibfield
  {journal} {\bibinfo  {journal} {\apj}\ }\textbf {\bibinfo {volume} {904}},\
  \bibinfo {eid} {47} (\bibinfo {year} {2020})}\BibitemShut {NoStop}%
\bibitem [{\citenamefont {{Held}}\ \emph {et~al.}(2024)\citenamefont {{Held}},
  \citenamefont {{Mamatsashvili}},\ and\ \citenamefont
  {{Pessah}}}]{2024MNRAS.530.2232H}%
  \BibitemOpen
  \bibfield  {author} {\bibinfo {author} {\bibfnamefont {L.~E.}\ \bibnamefont
  {{Held}}}, \bibinfo {author} {\bibfnamefont {G.}~\bibnamefont
  {{Mamatsashvili}}},\ and\ \bibinfo {author} {\bibfnamefont {M.~E.}\
  \bibnamefont {{Pessah}}},\ }\bibfield  {title} {\bibinfo {title} {{MRI
  turbulence in vertically stratified accretion discs at large magnetic Prandtl
  numbers}},\ }\href {https://doi.org/10.1093/mnras/stae929} {\bibfield
  {journal} {\bibinfo  {journal} {\mnras}\ }\textbf {\bibinfo {volume} {530}},\
  \bibinfo {pages} {2232} (\bibinfo {year} {2024})}\BibitemShut {NoStop}%
\bibitem [{\citenamefont {{Begelman}}\ and\ \citenamefont
  {{Armitage}}(2023)}]{2023MNRAS.521.5952B}%
  \BibitemOpen
  \bibfield  {author} {\bibinfo {author} {\bibfnamefont {M.~C.}\ \bibnamefont
  {{Begelman}}}\ and\ \bibinfo {author} {\bibfnamefont {P.~J.}\ \bibnamefont
  {{Armitage}}},\ }\bibfield  {title} {\bibinfo {title} {{Saturation of the
  magnetorotational instability and the origin of magnetically elevated
  accretion discs}},\ }\href {https://doi.org/10.1093/mnras/stad914} {\bibfield
   {journal} {\bibinfo  {journal} {\mnras}\ }\textbf {\bibinfo {volume}
  {521}},\ \bibinfo {pages} {5952} (\bibinfo {year} {2023})}\BibitemShut
  {NoStop}%
\bibitem [{\citenamefont {{Begelman}}(2024)}]{2024MNRAS.534.3144B}%
  \BibitemOpen
  \bibfield  {author} {\bibinfo {author} {\bibfnamefont {M.~C.}\ \bibnamefont
  {{Begelman}}},\ }\bibfield  {title} {\bibinfo {title} {{A simple model of
  globally magnetized accretion discs}},\ }\href
  {https://doi.org/10.1093/mnras/stae2305} {\bibfield  {journal} {\bibinfo
  {journal} {\mnras}\ }\textbf {\bibinfo {volume} {534}},\ \bibinfo {pages}
  {3144} (\bibinfo {year} {2024})}\BibitemShut {NoStop}%
\bibitem [{\citenamefont {{Ebrahimi}}\ \emph {et~al.}(2009)\citenamefont
  {{Ebrahimi}}, \citenamefont {{Prager}},\ and\ \citenamefont
  {{Schnack}}}]{ebrahimi2009saturation}%
  \BibitemOpen
  \bibfield  {author} {\bibinfo {author} {\bibfnamefont {F.}~\bibnamefont
  {{Ebrahimi}}}, \bibinfo {author} {\bibfnamefont {S.~C.}\ \bibnamefont
  {{Prager}}},\ and\ \bibinfo {author} {\bibfnamefont {D.~D.}\ \bibnamefont
  {{Schnack}}},\ }\bibfield  {title} {\bibinfo {title} {{Saturation of
  Magnetorotational Instability Through Magnetic Field Generation}},\ }\href
  {https://doi.org/10.1088/0004-637X/698/1/233} {\bibfield  {journal} {\bibinfo
   {journal} {\apj}\ }\textbf {\bibinfo {volume} {698}},\ \bibinfo {pages}
  {233} (\bibinfo {year} {2009})}\BibitemShut {NoStop}%
\bibitem [{\citenamefont {{Heinemann}}\ \emph {et~al.}(2011)\citenamefont
  {{Heinemann}}, \citenamefont {{McWilliams}},\ and\ \citenamefont
  {{Schekochihin}}}]{2011PhRvL.107y5004H}%
  \BibitemOpen
  \bibfield  {author} {\bibinfo {author} {\bibfnamefont {T.}~\bibnamefont
  {{Heinemann}}}, \bibinfo {author} {\bibfnamefont {J.~C.}\ \bibnamefont
  {{McWilliams}}},\ and\ \bibinfo {author} {\bibfnamefont {A.~A.}\ \bibnamefont
  {{Schekochihin}}},\ }\bibfield  {title} {\bibinfo {title} {{Large-Scale
  Magnetic Field Generation by Randomly Forced Shearing Waves}},\ }\href
  {https://doi.org/10.1103/PhysRevLett.107.255004} {\bibfield  {journal}
  {\bibinfo  {journal} {\prl}\ }\textbf {\bibinfo {volume} {107}},\ \bibinfo
  {eid} {255004} (\bibinfo {year} {2011})}\BibitemShut {NoStop}%
\bibitem [{\citenamefont {{Ebrahimi}}\ and\ \citenamefont
  {{Blackman}}(2016)}]{2016MNRAS.459.1422E}%
  \BibitemOpen
  \bibfield  {author} {\bibinfo {author} {\bibfnamefont {F.}~\bibnamefont
  {{Ebrahimi}}}\ and\ \bibinfo {author} {\bibfnamefont {E.~G.}\ \bibnamefont
  {{Blackman}}},\ }\bibfield  {title} {\bibinfo {title} {{Radially dependent
  large-scale dynamos in global cylindrical shear flows and the local cartesian
  limit}},\ }\href {https://doi.org/10.1093/mnras/stw724} {\bibfield  {journal}
  {\bibinfo  {journal} {\mnras}\ }\textbf {\bibinfo {volume} {459}},\ \bibinfo
  {pages} {1422} (\bibinfo {year} {2016})}\BibitemShut {NoStop}%
\bibitem [{\citenamefont {{Gressel}}\ and\ \citenamefont
  {{Pessah}}(2015)}]{2015ApJ...810...59G}%
  \BibitemOpen
  \bibfield  {author} {\bibinfo {author} {\bibfnamefont {O.}~\bibnamefont
  {{Gressel}}}\ and\ \bibinfo {author} {\bibfnamefont {M.~E.}\ \bibnamefont
  {{Pessah}}},\ }\bibfield  {title} {\bibinfo {title} {{Characterizing the
  Mean-field Dynamo in Turbulent Accretion Disks}},\ }\href
  {https://doi.org/10.1088/0004-637X/810/1/59} {\bibfield  {journal} {\bibinfo
  {journal} {\apj}\ }\textbf {\bibinfo {volume} {810}},\ \bibinfo {eid} {59}
  (\bibinfo {year} {2015})}\BibitemShut {NoStop}%
\bibitem [{\citenamefont {{Shi}}\ \emph {et~al.}(2016)\citenamefont {{Shi}},
  \citenamefont {{Stone}},\ and\ \citenamefont
  {{Huang}}}]{2016MNRAS.456.2273S}%
  \BibitemOpen
  \bibfield  {author} {\bibinfo {author} {\bibfnamefont {J.-M.}\ \bibnamefont
  {{Shi}}}, \bibinfo {author} {\bibfnamefont {J.~M.}\ \bibnamefont {{Stone}}},\
  and\ \bibinfo {author} {\bibfnamefont {C.~X.}\ \bibnamefont {{Huang}}},\
  }\bibfield  {title} {\bibinfo {title} {{Saturation of the magnetorotational
  instability in the unstratified shearing box with zero net flux: convergence
  in taller boxes}},\ }\href {https://doi.org/10.1093/mnras/stv2815} {\bibfield
   {journal} {\bibinfo  {journal} {\mnras}\ }\textbf {\bibinfo {volume}
  {456}},\ \bibinfo {pages} {2273} (\bibinfo {year} {2016})}\BibitemShut
  {NoStop}%
\bibitem [{\citenamefont {{Dhang}}\ \emph {et~al.}(2024)\citenamefont
  {{Dhang}}, \citenamefont {{Bendre}},\ and\ \citenamefont
  {{Subramanian}}}]{2024MNRAS.530.2778D}%
  \BibitemOpen
  \bibfield  {author} {\bibinfo {author} {\bibfnamefont {P.}~\bibnamefont
  {{Dhang}}}, \bibinfo {author} {\bibfnamefont {A.~B.}\ \bibnamefont
  {{Bendre}}},\ and\ \bibinfo {author} {\bibfnamefont {K.}~\bibnamefont
  {{Subramanian}}},\ }\bibfield  {title} {\bibinfo {title} {{Shedding light on
  the MRI-driven dynamo in a stratified shearing box}},\ }\href
  {https://doi.org/10.1093/mnras/stae1011} {\bibfield  {journal} {\bibinfo
  {journal} {\mnras}\ }\textbf {\bibinfo {volume} {530}},\ \bibinfo {pages}
  {2778} (\bibinfo {year} {2024})}\BibitemShut {NoStop}%
\bibitem [{\citenamefont {{Wissing}}\ \emph {et~al.}(2022)\citenamefont
  {{Wissing}}, \citenamefont {{Shen}}, \citenamefont {{Wadsley}},\ and\
  \citenamefont {{Quinn}}}]{2022A&A...659A..91W}%
  \BibitemOpen
  \bibfield  {author} {\bibinfo {author} {\bibfnamefont {R.}~\bibnamefont
  {{Wissing}}}, \bibinfo {author} {\bibfnamefont {S.}~\bibnamefont {{Shen}}},
  \bibinfo {author} {\bibfnamefont {J.}~\bibnamefont {{Wadsley}}},\ and\
  \bibinfo {author} {\bibfnamefont {T.}~\bibnamefont {{Quinn}}},\ }\bibfield
  {title} {\bibinfo {title} {{Magnetorotational instability with smoothed
  particle hydrodynamics}},\ }\href
  {https://doi.org/10.1051/0004-6361/202141206} {\bibfield  {journal} {\bibinfo
   {journal} {\aap}\ }\textbf {\bibinfo {volume} {659}},\ \bibinfo {eid} {A91}
  (\bibinfo {year} {2022})}\BibitemShut {NoStop}%
\bibitem [{\citenamefont {{Zier}}\ and\ \citenamefont
  {{Springel}}(2022)}]{2022MNRAS.517.2639Z}%
  \BibitemOpen
  \bibfield  {author} {\bibinfo {author} {\bibfnamefont {O.}~\bibnamefont
  {{Zier}}}\ and\ \bibinfo {author} {\bibfnamefont {V.}~\bibnamefont
  {{Springel}}},\ }\bibfield  {title} {\bibinfo {title} {{Simulating the
  magnetorotational instability on a moving mesh with the shearing box
  approximation}},\ }\href {https://doi.org/10.1093/mnras/stac2831} {\bibfield
  {journal} {\bibinfo  {journal} {\mnras}\ }\textbf {\bibinfo {volume} {517}},\
  \bibinfo {pages} {2639} (\bibinfo {year} {2022})}\BibitemShut {NoStop}%
\bibitem [{\citenamefont {{Skoutnev}}\ \emph {et~al.}(2022)\citenamefont
  {{Skoutnev}}, \citenamefont {{Squire}},\ and\ \citenamefont
  {{Bhattacharjee}}}]{Skoutnev+2022}%
  \BibitemOpen
  \bibfield  {author} {\bibinfo {author} {\bibfnamefont {V.}~\bibnamefont
  {{Skoutnev}}}, \bibinfo {author} {\bibfnamefont {J.}~\bibnamefont
  {{Squire}}},\ and\ \bibinfo {author} {\bibfnamefont {A.}~\bibnamefont
  {{Bhattacharjee}}},\ }\bibfield  {title} {\bibinfo {title} {{On large-scale
  dynamos with stable stratification and the application to stellar radiative
  zones}},\ }\href {https://doi.org/10.1093/mnras/stac2676} {\bibfield
  {journal} {\bibinfo  {journal} {\mnras}\ }\textbf {\bibinfo {volume} {517}},\
  \bibinfo {pages} {526} (\bibinfo {year} {2022})}\BibitemShut {NoStop}%
\bibitem [{\citenamefont {{R{\"a}dler}}(1986)}]{1986AN....307...89R}%
  \BibitemOpen
  \bibfield  {author} {\bibinfo {author} {\bibfnamefont {K.~H.}\ \bibnamefont
  {{R{\"a}dler}}},\ }\bibfield  {title} {\bibinfo {title} {{Investigations of
  spherical kinematic mean-field dynamo models}},\ }\href
  {https://doi.org/10.1002/asna.2113070205} {\bibfield  {journal} {\bibinfo
  {journal} {Astronomische Nachrichten}\ }\textbf {\bibinfo {volume} {307}},\
  \bibinfo {pages} {89} (\bibinfo {year} {1986})}\BibitemShut {NoStop}%
\bibitem [{\citenamefont {{Rogachevskii}}\ and\ \citenamefont
  {{Kleeorin}}(2003)}]{2003PhRvE..68c6301R}%
  \BibitemOpen
  \bibfield  {author} {\bibinfo {author} {\bibfnamefont {I.}~\bibnamefont
  {{Rogachevskii}}}\ and\ \bibinfo {author} {\bibfnamefont {N.}~\bibnamefont
  {{Kleeorin}}},\ }\bibfield  {title} {\bibinfo {title} {{Electromotive force
  and large-scale magnetic dynamo in a turbulent flow with a mean shear}},\
  }\href {https://doi.org/10.1103/PhysRevE.68.036301} {\bibfield  {journal}
  {\bibinfo  {journal} {\pre}\ }\textbf {\bibinfo {volume} {68}},\ \bibinfo
  {eid} {036301} (\bibinfo {year} {2003})}\BibitemShut {NoStop}%
\bibitem [{\citenamefont {{Rogachevskii}}\ and\ \citenamefont
  {{Kleeorin}}(2004)}]{2004PhRvE..70d6310R}%
  \BibitemOpen
  \bibfield  {author} {\bibinfo {author} {\bibfnamefont {I.}~\bibnamefont
  {{Rogachevskii}}}\ and\ \bibinfo {author} {\bibfnamefont {N.}~\bibnamefont
  {{Kleeorin}}},\ }\bibfield  {title} {\bibinfo {title} {{Nonlinear theory of a
  ``shear-current'' effect and mean-field magnetic dynamos}},\ }\href
  {https://doi.org/10.1103/PhysRevE.70.046310} {\bibfield  {journal} {\bibinfo
  {journal} {\pre}\ }\textbf {\bibinfo {volume} {70}},\ \bibinfo {eid} {046310}
  (\bibinfo {year} {2004})}\BibitemShut {NoStop}%
\bibitem [{\citenamefont {{Squire}}\ and\ \citenamefont
  {{Bhattacharjee}}(2015)}]{2015PhRvL.115q5003S}%
  \BibitemOpen
  \bibfield  {author} {\bibinfo {author} {\bibfnamefont {J.}~\bibnamefont
  {{Squire}}}\ and\ \bibinfo {author} {\bibfnamefont {A.}~\bibnamefont
  {{Bhattacharjee}}},\ }\bibfield  {title} {\bibinfo {title} {{Generation of
  Large-Scale Magnetic Fields by Small-Scale Dynamo in Shear Flows}},\ }\href
  {https://doi.org/10.1103/PhysRevLett.115.175003} {\bibfield  {journal}
  {\bibinfo  {journal} {\prl}\ }\textbf {\bibinfo {volume} {115}},\ \bibinfo
  {eid} {175003} (\bibinfo {year} {2015})}\BibitemShut {NoStop}%
\bibitem [{\citenamefont {{Zhou}}\ and\ \citenamefont
  {{Blackman}}(2021)}]{2021MNRAS.507.5732Z}%
  \BibitemOpen
  \bibfield  {author} {\bibinfo {author} {\bibfnamefont {H.}~\bibnamefont
  {{Zhou}}}\ and\ \bibinfo {author} {\bibfnamefont {E.~G.}\ \bibnamefont
  {{Blackman}}},\ }\bibfield  {title} {\bibinfo {title} {{On the shear-current
  effect: toward understanding why theories and simulations have mutually and
  separately conflicted}},\ }\href {https://doi.org/10.1093/mnras/stab2469}
  {\bibfield  {journal} {\bibinfo  {journal} {\mnras}\ }\textbf {\bibinfo
  {volume} {507}},\ \bibinfo {pages} {5732} (\bibinfo {year}
  {2021})}\BibitemShut {NoStop}%
\bibitem [{\citenamefont {{Squire}}\ and\ \citenamefont
  {{Bhattacharjee}}(2016)}]{2016JPlPh..82b5301S}%
  \BibitemOpen
  \bibfield  {author} {\bibinfo {author} {\bibfnamefont {J.}~\bibnamefont
  {{Squire}}}\ and\ \bibinfo {author} {\bibfnamefont {A.}~\bibnamefont
  {{Bhattacharjee}}},\ }\bibfield  {title} {\bibinfo {title} {{The magnetic
  shear-current effect: generation of large-scale magnetic fields by the
  small-scale dynamo}},\ }\href {https://doi.org/10.1017/S0022377816000258}
  {\bibfield  {journal} {\bibinfo  {journal} {Journal of Plasma Physics}\
  }\textbf {\bibinfo {volume} {82}},\ \bibinfo {eid} {535820201} (\bibinfo
  {year} {2016})}\BibitemShut {NoStop}%
\bibitem [{\citenamefont {{Brandenburg et al. (Pencil Code
  Collaboration)}}(2021)}]{2021JOSS....6.2807P}%
  \BibitemOpen
  \bibfield  {author} {\bibinfo {author} {\bibfnamefont {A.}~\bibnamefont
  {{Brandenburg et al. (Pencil Code Collaboration)}}},\ }\bibfield  {title}
  {\bibinfo {title} {{The Pencil Code, a modular MPI code for partial
  differential equations and particles: multipurpose and
  multiuser-maintained}},\ }\href {https://doi.org/10.21105/joss.02807}
  {\bibfield  {journal} {\bibinfo  {journal} {Journal of Open Source Software}\
  }\textbf {\bibinfo {volume} {6}},\ \bibinfo {eid} {2807} (\bibinfo {year}
  {2021})}\BibitemShut {NoStop}%
\bibitem [{\citenamefont {{Ebrahimi}}\ and\ \citenamefont
  {{Pharr}}(2022)}]{ebrahimi2022nonlocal}%
  \BibitemOpen
  \bibfield  {author} {\bibinfo {author} {\bibfnamefont {F.}~\bibnamefont
  {{Ebrahimi}}}\ and\ \bibinfo {author} {\bibfnamefont {M.}~\bibnamefont
  {{Pharr}}},\ }\bibfield  {title} {\bibinfo {title} {{A Nonlocal
  Magneto-curvature Instability in a Differentially Rotating Disk}},\ }\href
  {https://doi.org/10.3847/1538-4357/ac892d} {\bibfield  {journal} {\bibinfo
  {journal} {\apj}\ }\textbf {\bibinfo {volume} {936}},\ \bibinfo {eid} {145}
  (\bibinfo {year} {2022})}\BibitemShut {NoStop}%
\bibitem [{\citenamefont {{Ebrahimi}}\ and\ \citenamefont
  {{Haywood}}(2025)}]{ebrahimi2025generalized}%
  \BibitemOpen
  \bibfield  {author} {\bibinfo {author} {\bibfnamefont {F.}~\bibnamefont
  {{Ebrahimi}}}\ and\ \bibinfo {author} {\bibfnamefont {A.}~\bibnamefont
  {{Haywood}}},\ }\bibfield  {title} {\bibinfo {title} {{A generalized
  effective potential for differentially rotating plasmas}},\ }\href
  {https://doi.org/10.1063/5.0251633} {\bibfield  {journal} {\bibinfo
  {journal} {Physics of Plasmas}\ }\textbf {\bibinfo {volume} {32}},\ \bibinfo
  {eid} {030702} (\bibinfo {year} {2025})}\BibitemShut {NoStop}%
\bibitem [{\citenamefont {{Brandenburg}}\ and\ \citenamefont
  {{Sokoloff}}(2002)}]{2002GApFD..96..319B}%
  \BibitemOpen
  \bibfield  {author} {\bibinfo {author} {\bibfnamefont {A.}~\bibnamefont
  {{Brandenburg}}}\ and\ \bibinfo {author} {\bibfnamefont {D.}~\bibnamefont
  {{Sokoloff}}},\ }\bibfield  {title} {\bibinfo {title} {{Local and Nonlocal
  Magnetic Diffusion and Alpha-Effect Tensors in Shear Flow Turbulence}},\
  }\href {https://doi.org/10.1080/03091920290032974} {\bibfield  {journal}
  {\bibinfo  {journal} {Geophysical and Astrophysical Fluid Dynamics}\ }\textbf
  {\bibinfo {volume} {96}},\ \bibinfo {pages} {319} (\bibinfo {year}
  {2002})}\BibitemShut {NoStop}%
\end{thebibliography}%
	
	\clearpage
	
	\renewcommand{\appendixname}{}

	\onecolumngrid
	\begin{center}
		\textbf{\large Supplemental Material}
	\end{center}
	\twocolumngrid
	
	\setcounter{equation}{0}
	\setcounter{figure}{0}
	\setcounter{table}{0}
	\setcounter{section}{0}
	\renewcommand{\thesection}{S\arabic{section}}
	\renewcommand{\thefigure}{S\arabic{figure}}
	\renewcommand{\thetable}{S\arabic{table}}
	
	\counterwithout{equation}{section}
	\renewcommand{\theequation}{S\arabic{equation}}


\section{Numerical Setup}  \label{sec:method}

This study employs direct numerical simulations (DNS) of unstratified, zero-net-magnetic-flux MRI turbulence in a Keplerian accretion disk. We adopt a Cartesian shearing-sheet approximation, where differential rotation is locally represented as a linear shear flow, $\bec{U}^0 = -q \Omega x \hat{y} $, with a uniform rotation rate, $\bec{\Omega} = \Omega \hat{z}$, and a shear parameter of $q=3/2$. Here, $x$, $y$, and $z$ correspond to the radial, azimuthal, and vertical directions, respectively. The system evolves according to the standard magnetohydrodynamic (MHD) equations in the rotating shearing frame:
\begin{align}
	\frac{\mathcal{D} {\bec A} }{ \mathcal{D} t } =& q \Omega A_y \hat{x} + {\bec U} \times {\bec B} - \eta \mu_0 \bec J \;,
	\label{eq:induction} \\
	\frac{\mathcal{D} {\bec U} }{ \mathcal{D} t } =& - ({\bec U } \cdot \nabla )
	{\bec U} + q \Omega U_x \hat{y} - \frac{1}{\rho} \nabla P_g
	+ \frac{1}{\rho} {\bec J } \times {\bec B}  \nonumber \\
	& - 2 {\bec \Omega} \times {\bec U} 
	+ \frac{1}{\rho} \nabla \cdot 2\nu \rho \mathbf{S} \;,
	\label{eq:navier_stokes} \\
	\frac{\mathcal{D} {\ln \rho} }{ \mathcal{D} t } =& - ({\bec U} \cdot \nabla) \ln \rho - \nabla \cdot {\bec U} \;,
	\label{eq:continuity}
\end{align}
where $\bec U$ is the velocity field, $\rho$ the density, $P_g$ the thermal pressure, $\eta$ the magnetic diffusivity, and $\nu$ the microscopic viscosity. 
Here, $\mathcal{D}/\mathcal{D} t  =  \partial /\partial t - q\Omega x \partial_y$ includes advection by the background shear.
The rate-of-strain tensor is defined as $\mathbf{S}_{ij}=\frac{1}{2}(\bec U_{i,j}+\bec U_{j,i} - \frac{2}{3} \delta_{ij} \nabla \cdot \bec U )$. We assume an isothermal equation of state, $P_g = \rho c_s^2$, with constant sound speed $c_s$. We solve Eqs.~(\ref{eq:induction})--(\ref{eq:continuity}) using the \textsc{Pencil Code} \cite{2021JOSS....6.2807P}, a high-order finite-difference solver with sixth-order spatial accuracy and third-order time integration. The solenoidal constraint $\bm \nabla \cdot \bm{B} = 0$ is enforced by evolving the magnetic vector potential $\bm A$, where $\bm B = \bm \nabla \times \bm A$.

The computational domain is a cubic box spanning $-L_x/2 \leq x \leq L_x/2$, $-L_y/2 \leq y \leq L_y/2$, and $-L_z/2 \leq z \leq L_z/2$, with a uniform grid resolution of $256^3$. We impose periodic boundary conditions in the $y$ (azimuthal) and $z$ (vertical) directions, and shear-periodic conditions in the $x$ (radial) direction. All quantities are normalized such that length is scaled by box size $L$, velocity by $c_s$, density by its initial value $\rho_0$, and magnetic field by $(\mu_0 \rho_0 c_s^2)^{1/2}$, where we set $L = \Omega = \rho_0 = \mu_0 = c_s = 1$.

The initial velocity field is Gaussian random noise with an amplitude of $10^{-4} c_s$. The initial zero-net-flux vertical magnetic field is set as $\bec{B} = B_0 \sin (k_x x) \hat{z}$, corresponding to a vector potential $\bec{A} = A_0 \cos (k_x x) \hat{y}$, with $A_0 = 0.005$ and $k_x = 2\pi / L_x$. The plasma beta parameter is $\beta = 2\mu_0 P_g / \bar B_0^2 \simeq 1014$. From linear MRI analysis, the wavenumber of maximum growth is $k_{\text{max}} / k_1 = \sqrt{15/16} (\Omega / v_{A,0}) / k_1 \approx 5$, where $v_{A,0} = B_0 / \sqrt{\mu_0 \rho_0}$ is the initial Alfv\'{e}n speed, and $k_1 = 2\pi / L$. MRI-driven turbulence reaches a statistical steady state, with a root mean square (rms) velocity of $U_{\text{rms}} \sim 0.1 c_s$, indicating weak compressibility. The fluid and magnetic Reynolds numbers are defined as $\text{Re} \equiv U_{\text{rms}} L / \nu$ and $\text{Rm} \equiv U_{\text{rms}} L / \eta$. For our DNS runs, we set $\nu = 8 \times 10^{-5}$ and $\eta = 2 \times 10^{-5}$, yielding $\text{Rm} \simeq 5000$ and a magnetic Prandtl number of $\text{Pm} \equiv \text{Rm} / \text{Re} = 4$.

To confirm robustness, we perform additional simulations: (i) at $\text{Pm}=8$ ($\nu = 1.6 \times 10^{-4}$, $\eta = 2 \times 10^{-5}$) in a cubic domain with resolution $256^3$, and (ii) in a vertically elongated box with aspect ratio $L_x:L_y:L_z = 1:1:8$ and resolution $128 \times 128 \times 1024$ at $\text{Pm}=8$ ($\nu = 3.2 \times 10^{-4}$, $\eta = 4 \times 10^{-5}$). Both cases yield consistent nonhelical dynamo behavior, supporting the robustness of our conclusions. Unless otherwise stated, results in the main text correspond to the $\text{Pm}=4$ cubic box.

For the evaluation of turbulent correlations, we first compute horizontal planar averages to define the mean fields. Fluctuations are then obtained by subtracting these mean fields from the total fields and used to construct the relevant correlators. A second planar average is finally applied to obtain the EMF, stress tensors, and third-order correlators.

Linear MRI Dynamo Growth Rate:
	In the early, linear phase of the MRI, the evolution of the horizontally averaged mean magnetic field can be approximated by retaining only the dominant source term associated with the negative off-diagonal resistivity, $-\eta_{yx}$. All other terms (in Eq. 4) that act as sinks of magnetic energy are collectively modeled as a single effective diagonal turbulent resistivity, denoted by $\eta_t$. Assuming a mean-field mode with vertical structure $\propto \exp(\gamma t + i k_z z)$, the dispersion relation for the growth rate becomes $\gamma = k_z \sqrt{\eta_{yx} (-q\Omega + \eta_{xy} k_z^2)} - \eta_t k_z^2$. In the limit where $|q\Omega| \gg \eta_{xy} k_z^2$, this simplifies to $\gamma \simeq k_z \sqrt{ -q\Omega \eta_{yx}} - \eta_t k_z^2$. This form highlights that a negative $\eta_{yx}$ drives the exponential growth of the large-scale magnetic field, while $\eta_t$ represents the net damping effect from all dissipative processes.


\section{Pressure Fluctuations} \label{sec:pressure_fluctuations}

\begin{figure}
	\includegraphics[width=0.85\columnwidth]{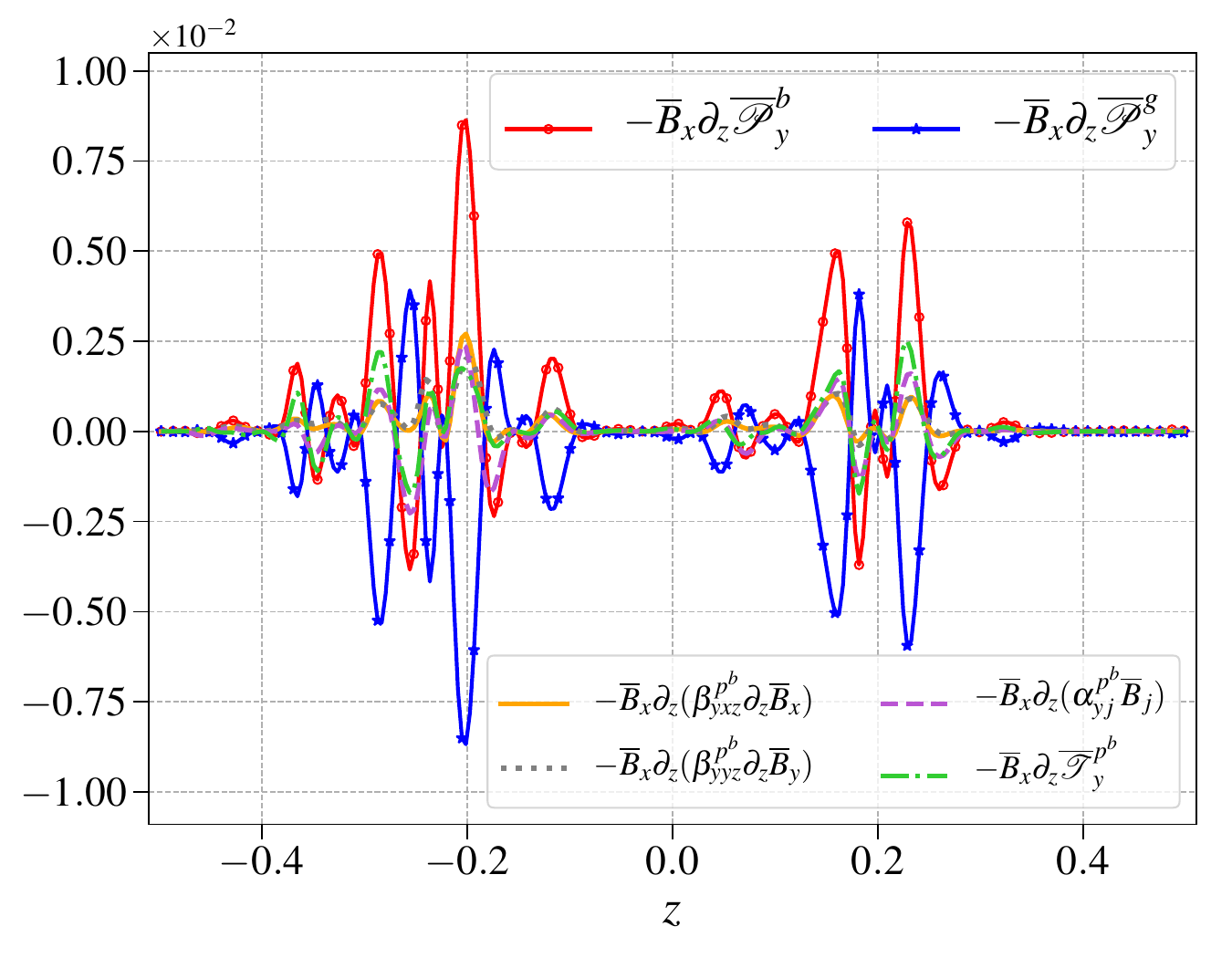}  
	\caption{	
		Spatial variation of terms associated with magnetic (Eq.~\ref{eq:P_b-y}) and gas pressure fluctuations (Eq.~\ref{eq:P_g-y}) contributing to $\bar B_x$ evolution via $\mathcal{\bar E}_y$ (Eq.~\ref{eq:meanBx_xy_2}) during the exponential growth phase of MRI, evaluated at $t/T_{\text{orb}} = 3.34$. The total effect of these fluctuations ($\mathcal{\bar P}_y$) is shown in \Fig{fig:BxdtBx_z_xyaver}(a), while individual contributions from four components of magnetic pressure fluctuations (Eq.~\ref{eq:P_b-y}) are displayed in the bottom legends.}
	\label{fig:pressure_z_xyaver_T3p34}
\end{figure}

We now analyze the contributions from gas and magnetic pressure fluctuations, and their effect on the mean EMF that drives large-scale radial magnetic fields. The pressure fluctuations become significant in the vertical component of the fluctuating Lorentz force, particularly in the evolution of $u_z$, which contributes to the EMFs through $\bar{F}_{zx} = \left< u_z b_x \right>$ and $\bar{F}_{zy} = \left< u_z b_y \right>$. Instead of detailing each term in the evolution equations for $\bar{F}_{zx}$ and $\bar{F}_{zy}$, we focus on the contributions from gas $(\mathcal{\bar P}^g_y)$ and magnetic $(\mathcal{\bar P}^b_y)$ pressure fluctuations, where ($\mathcal{\bar P}_y = \mathcal{\bar P}^g_y + \mathcal{\bar P}^b_y$), as they appear in $\mathcal{\bar E}_y$ (Eq.~\ref{eq:emfy}) and contribute to the evolution of $\bar B_{x}$ (Eq.~\ref{eq:meanBx_xy_2}), which is the primary focus of this study. The magnetic pressure fluctuations, given by $p_b = (b^2 + 2\bar B \cdot b - \bar{b^2} )/2\mu_0 $, yield
\begin{subequations}
	\begin{align}
		\mathcal{\bar P}^b_y = & \frac{-1}{q(2-q)\rho \Omega} \Bigg[ 
		(2-q) \left\{ \bar M_{yy} \partial_z \bar B_y + \bar M_{xy} \partial_z \bar B_x \right\} \nonumber\\ 
		& + \bar B_k \left\{ -q \frac{\langle b_z\partial_y b_k \rangle}{\mu_0} + (2-q) \frac{\langle b_y\partial_z b_k \rangle}{\mu_0} \right\} 
		\nonumber\\ 
		& 
		+ \left\{- q \frac{\langle b_z \partial_y b^2 \rangle}{2 \mu_0} + (2-q) \frac{\langle b_y \partial_z b^2 \rangle}{2 \mu_0} \right\} \Bigg], \nonumber\\ 
		& \hspace{-0.4cm} = \alpha_{yj}^{p^b} \bar B_j + \beta_{yyz}^{p^b} \partial_z \bar B_y + \beta_{yxz}^{p^b} \partial_z \bar B_x + \mathcal{\bar T}_y^{p^b} ,
		\label{eq:P_b-y}
	\end{align}
	\begin{equation}
		\mathcal{\bar P}^g_y = \frac{-1}{q(2-q)\rho \Omega} \Big\{- q \langle b_z\partial_y p_g\rangle + (2-q) \langle b_y\partial_z p_g\rangle \Big\} .
		\label{eq:P_g-y}
	\end{equation}
	\label{eq:P_t-y}
\end{subequations}
Magnetic pressure fluctuations contribute to the EMF through turbulent diffusivity tensors $\eta_{ij}$, $\alpha_{ij}$ tensors, and third-order correlators. A crucial aspect is the negative $\eta_{ij}$ tensors, which can drive dynamo action, given by
\begin{equation}
	\eta^{p^b}_{yx} = \beta_{yyz}^{p^b} = - \frac{1}{\rho q \Omega} \bar M_{yy}; \ \eta^{p^b}_{yy} = - \beta_{yxz}^{p^b} = \frac{1}{\rho q \Omega} \bar M_{xy}.
	\label{eq:eta_pb}
\end{equation}
In MRI turbulence, $\bar M_{xy}$ is consistently negative, while $\bar M_{yy}$ remains positive. Consequently, both $\eta^{p^b}_{yx}$ and $\eta^{p^b}_{yy}$ are negative, serving as source terms for $\bar B_x$ generation. Previous studies on shear-flow turbulence reported negative $\eta_{yy}$, though without a detailed explanation, often deeming it unphysical \cite{2002GApFD..96..319B, 2016JPlPh..82b5301S, 2024MNRAS.530.2778D}. Here, we demonstrate that the negative $\eta_{yy}$ arises from magnetic pressure fluctuations via the correlator $\bar M_{xy}$, which is responsible for outward angular momentum transport in accretion disks. Despite their dynamo role, the contributions of magnetic pressure fluctuations are largely canceled by those from gas pressure fluctuations, resulting in a negligible net effect. These findings are illustrated in \Fig{fig:pressure_z_xyaver_T3p34}, which presents the contributions of gas and magnetic pressure fluctuations to $\mathcal{\bar E}_y$ during the exponential growth phase of MRI at $t/T_{\text{orb}} = 3.34$. The total contribution of these fluctuations ($\mathcal{\bar P}_y$) is shown in \Fig{fig:BxdtBx_z_xyaver}(a), while individual contributions from four distinct components of magnetic pressure fluctuations are displayed in the bottom legend.

\end{document}